\documentclass[aps,prf,preprint,longbibliography,superscriptaddress]{revtex4-1}
\usepackage{amsmath}
\usepackage{amssymb}
\usepackage{amsfonts}

\usepackage{color}
\usepackage{graphicx}
\usepackage{bm}

\newcommand{\new}{\color{black}}

\begin{document}

\title{Sedimentation of a rigid helix in viscous media}

\author{Martina Palusa}
\affiliation{SUPA, School of Physics and Astronomy, The University of Edinburgh, James Clerk Maxwell Building, Peter Guthrie Tait Road, Edinburgh, EH9 3FD, United Kingdom}

\author{Joost de Graaf}
\affiliation{Institute for Theoretical Physics, Center for Extreme Matter and Emergent Phenomena, Utrecht University, Princetonplein 5, 3584 CC Utrecht, The Netherlands}

\author{Aidan Brown}
\affiliation{SUPA, School of Physics and Astronomy, The University of Edinburgh, James Clerk Maxwell Building, Peter Guthrie Tait Road, Edinburgh, EH9 3FD, United Kingdom}

\author{Alexander Morozov}
\email{alexander.morozov@ed.ac.uk}
\affiliation{SUPA, School of Physics and Astronomy, The University of Edinburgh, James Clerk Maxwell Building, Peter Guthrie Tait Road, Edinburgh, EH9 3FD, United Kingdom}

\date{\today}

\begin{abstract}
We consider sedimentation of a rigid helical filament in a viscous fluid under gravity. In the Stokes limit, the drag forces and torques on the filament are approximated within the resistive-force theory. We develop an analytic approximation to the exact equations of motion that works well in the limit of a sufficiently large number of turns in the helix (larger than two, typically). For a wide range of initial conditions, our approximation predicts that the centre of the helix itself follows a helical path with the symmetry axis of the trajectory being parallel to the direction of gravity. The radius and the pitch of the trajectory scale as non-trivial powers of the number of turns in the original helix. For the initial conditions corresponding to an almost horizontal orientation of the helix, we predict trajectories that are either attracted towards the horizontal orientation, in which case the helix sediments in a straight line along the direction of gravity, or trajectories that form a helical-like path with many temporal frequencies involved. Our results provide new insight into the sedimentation of chiral objects and might be used to develop new techniques for their spatial separation.
\end{abstract}

\maketitle

\section{Introduction}

The microhydrodynamics of helices is emerging as an important research topic across many disciplines. Motivated by their abundance in microscopic organisms \cite{Lighthill73,Purcell77,BergBook,LaugaPowers2009}, helical shapes were recently studied in the context of self-propulsion \cite{LaugaPowers2009}, fabrication of magnetically-driven micro- and nano-scale robots \cite{Zhang2009,Fischer2011,Tottori2012,Peyer2013}, and soft microflow sensors \cite{Attia2009,Pham2015}. In a more general context, recent experimental and theoretical work suggests that chiral objects often follow helical trajectories when sedimenting in viscous fluids due to gravity \cite{Makino2005,Krapf2009,Tozzi2011}, while both chiral and non-chiral objects exhibit spatial drift under shear flow \cite{Makino2005a,Marcos2009,Wang2012}. The underlying physics of the latter phenomenon, which is at the origin of swimming bacteria assuming a particular orientation with respect to an external flow (the so-called \emph{rheotaxis}) \cite{Marcos2012}, was proposed as a means of spatial separation of chiral objects in viscous media by external electric fields or shear \cite{Makino2016,Makino2017}.

In this work, we study the dynamics of a rigid helical filament sedimenting in a viscous fluid under Stokes flow conditions \cite{HappelBrenner,KimKarrila}. Although previous work suggests that the helix is expected to follow a helical path with the symmetry axis of the trajectory being parallel to the direction of gravity, there are no simple analytical predictions connecting the geometrical parameters of the helix to the properties of its spatial trajectory and sedimentation speed that can be readily compared against potential experimental data. To this end, we employ the resistive-force theory \cite{Lighthill73} and 
study the orientational and positional dynamics of a sedimenting helical filament. Although the resistive-force theory fails not only quantitatively, but also qualitatively, for rather compact objects \cite{Pak2012}, 
it is reasonably successful for sufficiently loose coils \cite{Jung2007,Rodenborn2013,Koens2014}.
We consider long filaments and develop an analytic approximation that allows us to predict the spatial kinematics of sedimenting helices. We show that besides the helical trajectories reported previously \cite{Makino2005,Krapf2009,Tozzi2011}, there are other types of orbit available to a helix depending on the initial conditions, and we determine their basins of attraction.

\section{Problem setup}

\subsection{Kinematics}

\begin{figure}[h]
\includegraphics[width=0.6\textwidth]{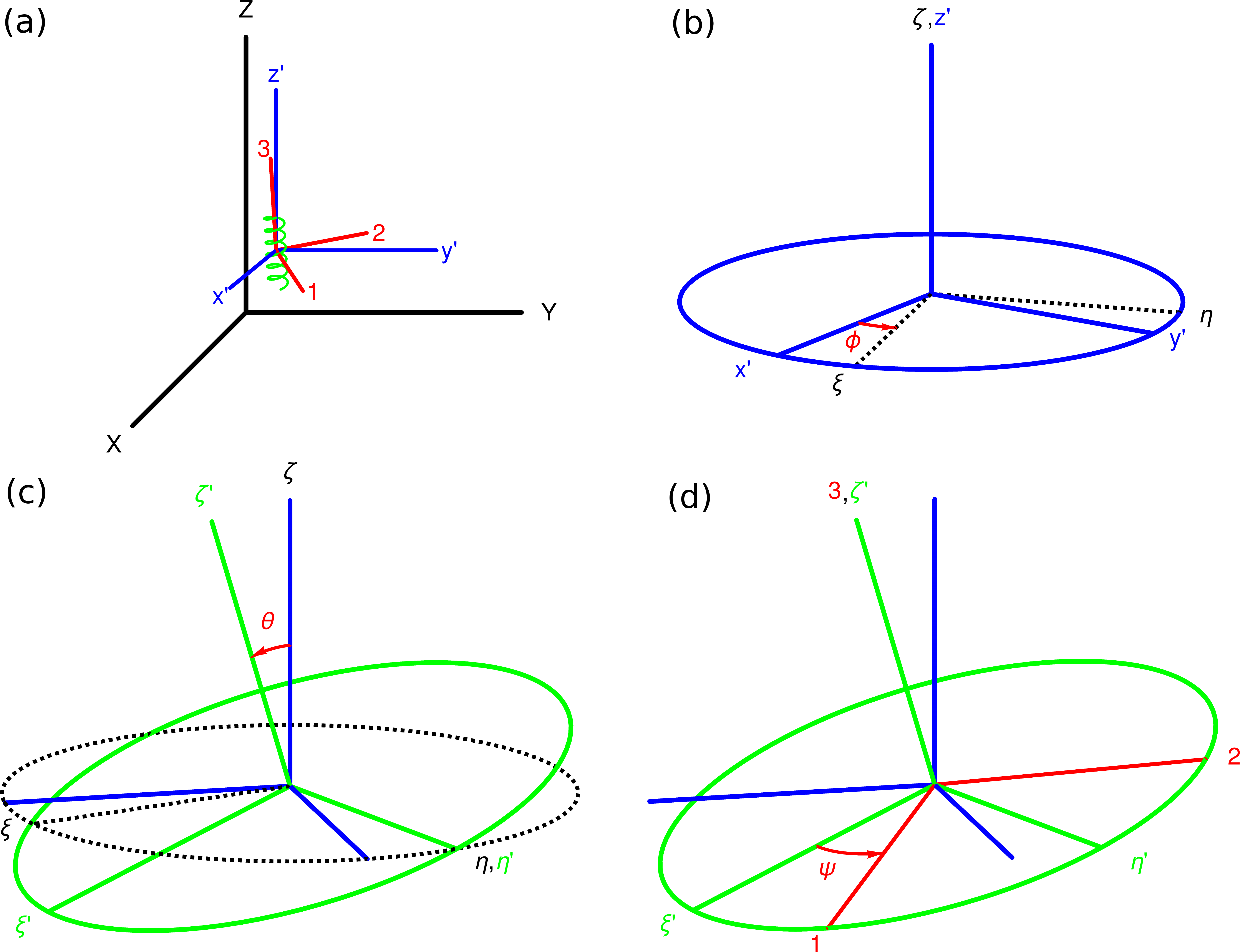}
\caption{(a) Definitions of the coordinate systems $\{X,Y,Z\}$, $\{x',y',z'\}$ and $\{1,2,3\}$. (b)-(d) Definition of the Euler angles used in this work.}
\label{Fig:DefEulAng}
\end{figure}

To describe the motion of a rigid helix, we introduce three Cartesian coordinate systems, see Fig.\ref{Fig:DefEulAng}(a). The first is the lab frame, $\{X,Y,Z\}$, with its $Z$-axis pointing in the direction opposite to the direction of gravity. The second frame, $\{x',y',z'\}$, is obtained by translating the lab frame with the helix; its axes are always parallel to the lab frame. Finally, the body frame, $\{1,2,3\}$, chosen along the principle directions of the helix (see below), rotates with the helix with respect to the $\{x',y',z'\}$ frame.

The body frame is related to the co-moving lab frame by three rotations that we define with the help of the Euler angles $\left(\theta,\phi,\psi\right)$: first the $\{x',y',z'\}$ frame is rotated around its $z'$-direction by the angle $\phi$, the resulting frame is  rotated around its $2$-axis by $\theta$, and, finally, that frame is rotated around its $3$-axis by the angle $\psi$. Note that our definition of the Euler angles differs from the commonly used convention given in Goldstein \cite{Goldstein}, see Fig.\ref{Fig:DefEulAng}(b)-(d). The Cartesian coordinates of a vector $\bm a$ in the co-moving lab frame and in the body frame are then related by the rotation matrix
\begin{align}
\left(a_{1},a_{2},a_{3}\right)^T = {\bm D}\cdot \left(a_{x'},a_{y'},a_{z'}\right)^T,
\label{TransformRule}
\end{align}
where $T$ denotes the transpose, and
\begin{align}
&{\bm D}(\phi, \theta, \psi) = \nonumber \\
 	&\begin{pmatrix}
	\cos\phi \cos\theta \cos\psi  - \sin\phi \sin\psi  & \quad \sin\phi \cos\theta \cos\psi  + \cos\phi \sin\psi  & \quad -\sin\theta \cos\psi  \\
	-\cos\phi \cos\theta \sin\psi  - \sin\phi \cos\psi  & \quad-\sin\phi \cos\theta \sin\psi  + \cos\phi \cos\psi  & \quad \sin\theta \sin\psi  \\
	\cos\phi \sin\theta  & \quad \sin\phi \sin\theta  & \quad \cos\theta 
	\end{pmatrix}.
\label{RotMatrix}
\end{align}
Finally, the angular velocities in the {\new lab frame and in the frame instantaneously coincident with the body frame} are related through the Euler angles
\begin{align}
&\Omega_{x'} = -\dot\theta \sin{\phi} + \dot\psi \cos{\phi} \sin{\theta}, \label{Ox} \\
&\Omega_{y'} = \dot\theta \cos{\phi} + \dot\psi \sin{\phi} \sin{\theta},  \label{Oy} \\
&\Omega_{z'} = \dot\phi +\dot\psi \cos{\theta}, \label{Oz}
\end{align}
and
\begin{align}
&\Omega_{1} = \dot\theta \sin{\psi} - \dot\phi \cos{\psi} \sin{\theta}, \label{O1} \\
&\Omega_{2} = \dot\theta \cos{\psi} + \dot\phi \sin{\psi} \sin{\theta}, \label{O2} \\
&\Omega_{3} = \dot\phi \cos{\theta} +\dot\psi, \label{O3}
\end{align}
where dot denotes the time derivative. 

\begin{figure}[h]
\includegraphics[width=0.45\textwidth]{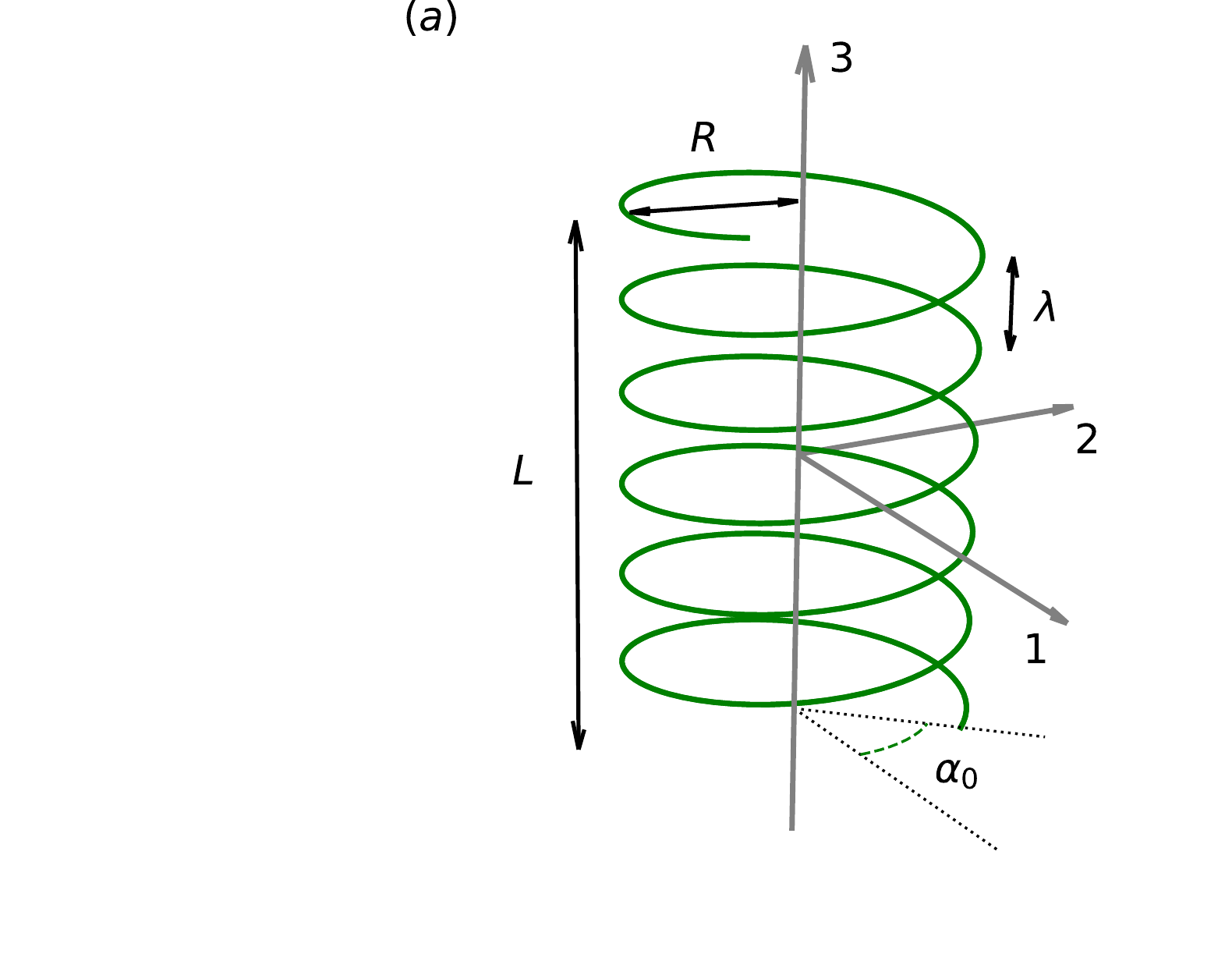}
\hspace{1cm}
\includegraphics[width=0.355\textwidth]{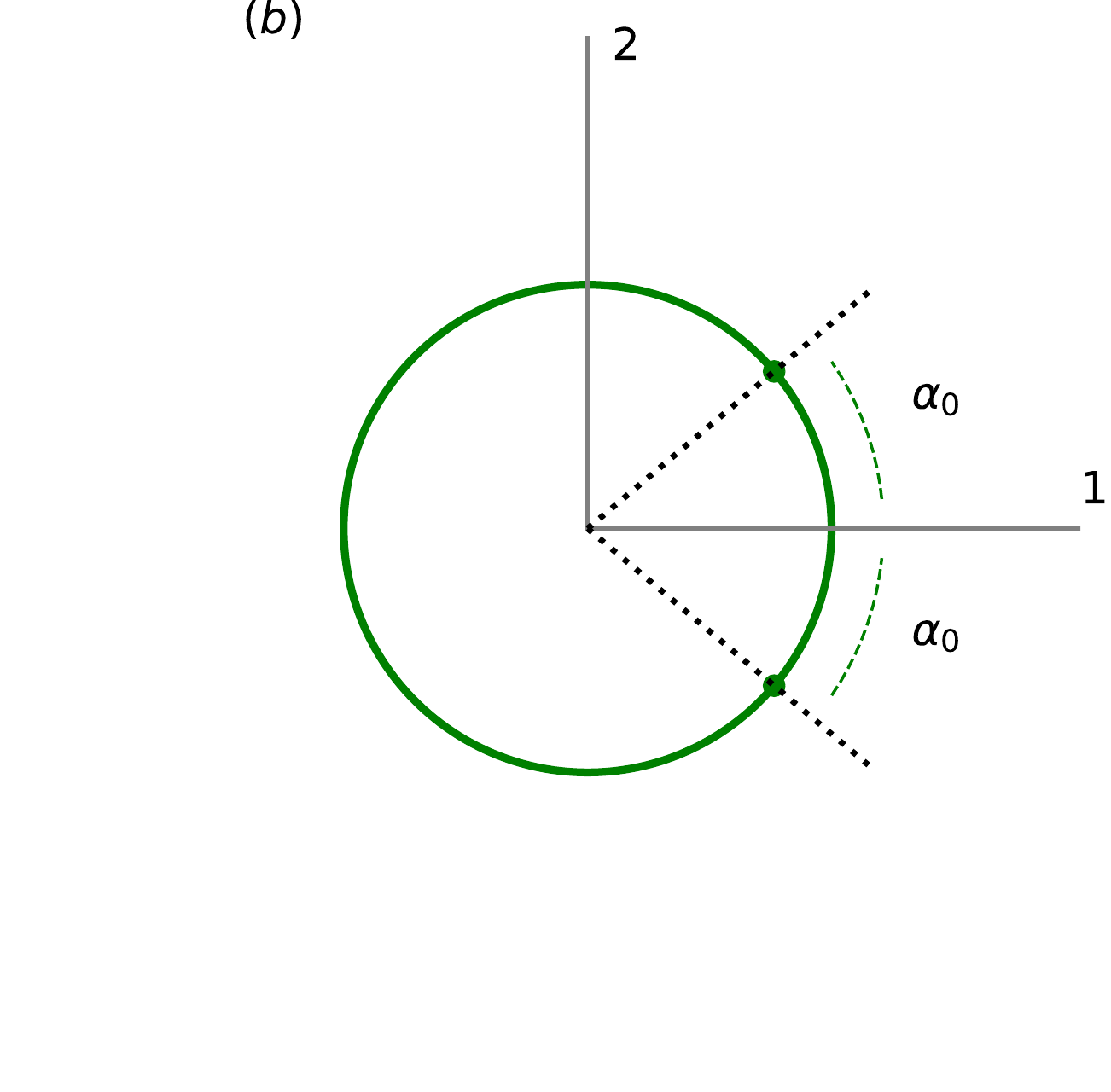}
\caption{Geometry of the helix: $R$ and $\lambda$ are the radius and the pitch, correspondingly, while $L$ is the total length of the helix along its symmetry axis; $\alpha_0$ is defined as the angle between the shortest distance between the end-point of the helix and its axis of symmetry, and the line obtained by rotation of the former in the $12$-plane until it is parallel to the $1$-direction. It serves as a measure of how much the number of helical turns, defined formally as $L/\lambda$, deviates from an integer (see the main text). (a) Side view. (b) Top view.}
\label{Fig:definitions}
\end{figure}

\subsection{Forces and torques on the helix in the body frame}

Here we calculate the body-frame forces and the torques acting on the helix moving through a viscous fluid. We assume that the helical filament is rigid and slender, and that the helix is sufficiently extended, and employ the resistive-force theory to approximate local hydrodynamic forces acting on small sections of the filament. We choose the helical axis to be oriented along the $3$-direction of the body frame. The position $\bm r$ of a point on the helical filament is parametrised by an angle $\alpha$ so that 
\begin{align}
{\bm r}_h(\alpha) = \left(R \cos{\alpha}, R \sin{\alpha}, \frac{\lambda}{2\pi}\alpha \right),
\label{rh}
\end{align}
where $R$ is the radius of the helix and $\lambda$ is its pitch; see Fig.\ref{Fig:definitions} for details. The angle $\alpha$ takes its values from the interval $\left[-\pi N+\alpha_0, \pi N-\alpha_0\right]$, where $\alpha_0$ determines by how much the helix deviates from having an integers number of turns $N$; we only consider $\alpha_0\in\left[-\pi/2,\pi/2\right)$ since any other value of $\alpha_0$ can be mapped onto this interval and a different value of $N$. The vertical extend of the helix, $L$, is determined from Eq.\eqref{rh} to read $L=\lambda\left(N-\alpha_0/\pi\right)$. In the following it will be more convenient to introduce the angle $\chi$ that the helical filament makes with the vertical axis and express the radius of the helix in terms of that angle: $R=\left(\lambda/2\pi\right)\tan{\chi}$. We, therefore, describe the helix by three parameters: its pitch $\lambda$, the angle $\chi$, and the total vertical length $L$.

Equation \eqref{rh} implicitly defines the axes of the body frame. For an integer number of turns, $\alpha_0=0$, the first axis points outwards along the shortest line drawn from the axis of the helix to one of its ends, while for $\alpha_0\ne0$, it points along the bisector of the angle formed by the corresponding lines drawn to both ends of the helix. The second axis is chosen so that the axes form a right-handed triplet. The origin of the body frame thus selected will be referred to as the centre of the helix, which coincide with its centre of mass for $\alpha_0=0$.

Within the resistive-force theory, the drag force acting on a small element $ds$ of the helical filament can be approximated by
\begin{align}
\bm{dF} = - \bigg[ K_{\parallel} \left({\bm V}\cdot {\bm t} \right){\bm t} + K_{\perp} \bigg\{ {\bm V}-\left({\bm V}\cdot {\bm t} \right){\bm t}\bigg\} \bigg] ds,
\label{dF}
\end{align}
where ${\bm V}$ is the velocity of the element relative to the surrounding fluid, and ${\bm t}$ is a unit vector along the direction of the filament at that point. For the curve ${\bm r}_h(\alpha)$, ${\bm t}$ is along the tangent $\partial{\bm r}_h(\alpha)/\partial \alpha$, i.e.  ${\bm t}=\left(-\sin{\alpha}\sin{\chi}, \cos{\alpha}\sin{\chi},\cos{\chi}\right)$, and $ds=(\lambda/2\pi)\,d\alpha/\cos{\chi}$.
To first approximation, the friction coefficients $K_{\parallel}$ and $K_{\perp}$ can be taken to be the drag coefficients of a straight rod \cite{HappelBrenner} moving parallel and perpendicular to its axis, correspondingly. While taking into account hydrodynamic interactions between the points of the same element of length $ds$,  this approach neglects interactions between adjacent elements of the filament. To effectively incorporate these interactions, Lighthill proposed the following expressions for the friction coefficients
\begin{align}
K_{\parallel} = \frac{4\pi\mu}{2\ln{(c_L \lambda/r)}-1}, \label{Kpar}\\
K_{\perp} = \frac{8\pi\mu}{2\ln{(c_L \lambda/r)}+1}, \label{Kperp}
\end{align}
where $\mu$ is the viscosity of the fluid, $r$ is the radius of the filament, and $c_L=0.18$ is the Lighthill constant; we also introduce $\gamma = K_\perp/K_\parallel$. In the following, when comparing our analytical results to the numerical solutions of the equations of motion in dimensionless form, we will set $\gamma=2$ to avoid introducing the radius of the filament $r$ as an additional parameter. When analysing our predictions in physical units, we will use Eqs.\eqref{Kpar} and \eqref{Kperp} for the friction coefficients in Eq.\eqref{dF}.

The total drag force acting on the helix is given by
\begin{align}
{\bm F} =\int_{-\pi N+\alpha_0}^{\pi N-\alpha_0} \bm{dF}(\alpha),
\label{Fint}
\end{align}
with $\bm{dF}(\alpha)$ from Eq.\eqref{dF}. In a similar fashion, the total torque applied to the helix by the drag forces calculated with respect to the origin of the body frame is given by
\begin{align}
{\bm T} =\int_{-\pi N+\alpha_0}^{\pi N-\alpha_0} \bm{r}_h(\alpha)\times \bm{dF}(\alpha).
\label{Tint}
\end{align}
We assume that the origin of the body frame moves through the fluid with the velocity $\bm{U}$ and that the helix is rotating with the angular velocity $\bm{\Omega}$, and, thus, the velocity of the point $\bm{r}_h(\alpha)$ on the filament  equals $\bm{V}(\alpha)=\bm{U} + \bm{\Omega}\times\bm{r}_h(\alpha)$. Substituting this expression into Eqs.\eqref{Fint} and \eqref{Tint} and performing integration over $\alpha$ yields
\begin{align}
\label{F1}
F_1 &=\frac{K_\parallel L}{\cos\chi} \bigg[ -\gamma U_1  \nonumber \\ 
&\qquad\quad + \frac{\gamma-1}{2}\left\{ (1+\Delta_1) U_1\sin^2\chi 
 -\left(\frac{1}{2}\cos{2\alpha_0} + 1 + \frac{3}{2}\Delta_1 \right)\frac{\lambda}{2\pi}\Omega_1 \sin^2\chi \right\}  \bigg], \\
\label{F2}
F_2 &= \frac{K_\parallel L}{\cos\chi} \left[ -\gamma\left( U_2-\Delta_2\frac{\lambda}{2\pi}\Omega_3 \tan\chi\right) + \frac{\gamma-1}{2}\bigg\{ (1-\Delta_1)U_2\sin^2\chi - \Delta_2 U_3 \sin2\chi \right.\nonumber \\
&  \left. \qquad \quad + \left(\frac{1}{2}\cos{2\alpha_0} - 1 + \frac{3}{2}\Delta_1 \right)\frac{\lambda}{2\pi}\Omega_2 \sin^2\chi - 2\Delta_2 \frac{\lambda}{2\pi} \Omega_3 \sin^2\chi \tan\chi \bigg\} \right],\\
\label{F3}
F_3 & = \frac{K_\parallel L}{\cos\chi} \left[ -\gamma\left( U_3+\Delta_2\frac{\lambda}{2\pi}\Omega_2 \tan\chi\right) + \left(\gamma-1\right)\bigg\{ -\Delta_2 U_2\sin\chi\cos\chi + U_3 \cos^2\chi \right.\nonumber \\
&  \left. \qquad \quad + \left((-1)^N\cos\alpha_0 + 2\Delta_2 \right)\frac{\lambda}{2\pi}\Omega_2 \sin\chi\cos\chi + \frac{\lambda}{2\pi}\Omega_3 \sin^2\chi \bigg\} \right],
\end{align}
and
\begin{align}
\label{T1}
T_1 &= \frac{\lambda}{4\pi}\frac{K_\parallel L}{\cos\chi} \bigg[ 
-\gamma\left( (1+\Delta_1)\tan^2\chi +\frac{2}{3}\left(\frac{\pi L}{\lambda}\right)^2\right)\frac{\lambda}{2\pi}\Omega_1  \nonumber \\
& \qquad\qquad - (\gamma-1)\left(\frac{1}{2}\cos{2\alpha_0} + 1 + \frac{3}{2}\Delta_1 \right) U_1 \sin^2\chi \nonumber \\
& \qquad\qquad + (\gamma-1)\left\{ 1+\frac{3}{2}\cos2\alpha_0 +\frac{5}{2}\Delta_1 + \left(\frac{\pi L}{\lambda}\right)^2\left(\frac{1}{3}-\Delta_1\right) \right\}\frac{\lambda}{2\pi}\Omega_1 \sin^2\chi \biggr], \\
\label{T2}
T_2 &= \frac{\lambda}{2\pi}\frac{K_\parallel L}{\cos\chi} \bigg[ 
-\gamma \Delta_2 U_3 \tan\chi - \frac{1}{2}\gamma\left( (1-\Delta_1)\tan^2\chi +\frac{2}{3}\left(\frac{\pi L}{\lambda}\right)^2\right)\frac{\lambda}{2\pi}\Omega_2  \nonumber \\
& \quad - \gamma\left( (-1)^N\cos\alpha_0 + \Delta_2 \right) \frac{\lambda}{2\pi}\Omega_3 \tan\chi + \frac{\gamma-1}{2}\left(\frac{1}{2}\cos{2\alpha_0} - 1 + \frac{3}{2}\Delta_1 \right) U_2 \sin^2\chi  \nonumber \\
& \qquad\qquad + (\gamma-1)\left( (-1)^N\cos\alpha_0 + 2\Delta_2 \right)\left(U_3+\frac{\lambda}{2\pi}\Omega_3 \tan^2\chi\right)\sin\chi\cos\chi \nonumber \\
&  \qquad\qquad + \frac{\gamma-1}{2}\left\{ 1-\frac{3}{2}\cos2\alpha_0 -\frac{5}{2}\Delta_1 + \left(\frac{\pi L}{\lambda}\right)^2\left(\frac{1}{3}+\Delta_1\right) \right\}\frac{\lambda}{2\pi}\Omega_2 \sin^2\chi 
\bigg], \\
\label{T3}
T_3 &= \frac{\lambda}{2\pi}\frac{K_\parallel L}{\cos\chi} \bigg[ 
\left( \sin^2\chi + \gamma\cos^2\chi \right)\Delta_2 U_2 \tan\chi \nonumber \\
& \qquad\qquad + (\gamma-1)U_3\sin^2\chi - \left( \sin^2\chi + \gamma\cos^2\chi \right) \frac{\lambda}{2\pi}\Omega_3 \tan^2\chi
 \nonumber \\
& \quad + \bigg\{ \left( \sin^2\chi + \gamma\cos^2\chi \right) (-1)^{N+1}\cos\alpha_0 + \left( 2(\gamma-1)\sin^2\chi - \gamma \right)\Delta_2 \bigg\}\frac{\lambda}{2\pi}\Omega_2 \tan\chi 
\bigg],
\end{align}
where we have introduced
\begin{align}
&\Delta_1 = \frac{\lambda}{2\pi L}\sin{2\alpha_0}, \nonumber\\
&\Delta_2 = \frac{(-1)^N\lambda}{\pi L}\sin{\alpha_0}. \nonumber
\end{align}
These expressions are similar to the ones obtained by Keller and Rubinow \cite{Keller1976} for a somewhat different definition of their helical geometry.

\subsection{Equations of motion}

In the Stokes limit, the motion of the helix is determined by the requirement that the total forces and torques acting on the helix vanish
\begin{align}
& \bm{F} + \bm{F}_g = 0, \\
&\bm{T} = 0,
\end{align}
where the gravity force is given by
\begin{align}
{\mathbf F_g} = -\left( 1 - \frac{\rho_f}{\rho_h}\right) M g\,\hat{\mathbf Z} \equiv - P \,\hat{\mathbf Z}.
\end{align}
Here, $M=\rho_h \pi r^2 L/\cos\chi$ is the mass of the helix, $\rho_h$ and $\rho_f$ are the densities of the helix and the suspending fluid, respectively, $g$ is the acceleration due to gravity, and $\hat{\mathbf Z}$ is a unit vector along the $Z$-direction of the lab frame. Transforming the force and torque balance to the body frame, the equations of motion read
\begin{align}
& F_1 + P\cos\psi\sin\theta = 0, \label{eom1}\\
& F_2 - P\sin\psi\sin\theta = 0, \\
& F_3 - P\cos\theta = 0,\\
& T_1 = T_2 = T_3 = 0. \label{eom4}
\end{align}
where the drag forces and torques are given by Eqs.\eqref{F1} - \eqref{T3}. In what follows, we adopt the solution methodology used in solid body mechanics \cite{Goldstein}. First, we solve the equations of motion to find the velocity and angular velocity components in the body frame, and use them to solve for the dynamics of the Euler angles. Finally, we use the time-dependent Euler angles to relate the body-frame kinematics to the lab-frame trajectories.

\section{Approximate solution for long helices}

The solution strategy outlined above is rather straightforward as the equations of motion, Eqs.\eqref{eom1}-\eqref{eom4}, are linear in the velocity and angular velocity components and can be easily solved. However, the large number of terms in Eqs.\eqref{F1} - \eqref{T3} results in rather cumbersome expressions that do not lead to further insight. Instead, we consider here the limit of long helices, which lends itself
to simple analytic treatment. We will demonstrate that the approximate solution thus developed is remarkably accurate even for relatively short helices.

To start, we rewrite the equations of motion, Eqs.\eqref{eom1}-\eqref{eom4}, in a dimensionless form by scaling all lengths with $L$ and time with the timescale $\tau$ defined as 
\begin{align}
\tau = \frac{K_\parallel \lambda^2}{P}\frac{ \pi c_0^2}{6(\gamma-1)\sin^2\chi \cos{\chi}},
\end{align}
where $c_0=\gamma + \gamma\cos^2{\chi} + \sin^2{\chi}$. This choice of timescale is not obvious and is made to simplify the dimensionless equations of motion in what follows. Dimensionless variables are denoted by a tilde.  

Next, we introduce $\epsilon=\lambda/L$ that we use as a small parameter. Analysis of the equations of motion shows that, to lowest order, all dimensionless velocity components are $O(\epsilon^2)$, $\tilde\Omega_1$ and $\tilde\Omega_2$ are $O(\epsilon^3)$, while $\tilde\Omega_3$ is $O(\epsilon)$. Therefore, up to $O(\epsilon^3)$, the angular velocities are given by
\begin{align}
&\tilde{\Omega}_1 = -\epsilon^3 \left(2+\cos{2\alpha_0} \right)\cos\psi\sin\theta, \\
&\tilde{\Omega}_2 =\epsilon^3 \left(2-\cos{2\alpha_0} \right)\sin\psi\sin\theta, \\
&\tilde{\Omega}_3 = -\epsilon \frac{\pi^2}{3\gamma}\frac{c_0^2}{\tan^2{\chi}}\cos{\theta} - \epsilon^2 \frac{2(-1)^N \pi c_0}{3(\gamma-1)}\frac{\sin{\alpha_0}\cos{\chi}}{\sin^3{\chi}}\sin{\psi}\sin{\theta} \\
& \qquad\quad - \epsilon^3 \frac{(-1)^N}{3\tan{\chi}}\left( 4\cos{\alpha_0} - \cos{3\alpha_0}\right)\sin{\psi}\sin{\theta}.
\end{align}
Using Eqs.\eqref{O1}-\eqref{O3}, we obtain for the Euler angles
\begin{align}
& \frac{\partial \theta}{\partial \tilde{t}} = -\epsilon^3 \cos{2\alpha_0}\sin{2\psi}\sin{\theta}, 
\label{eqtheta}\\
& \frac{\partial \phi}{\partial \tilde{t}} = \epsilon^3 \left( 2 + \cos{2\alpha_0} \cos{2\psi} \right), 
\label{eqphi}\\
& \frac{\partial \psi}{\partial \tilde{t}} = -\epsilon \frac{\pi^2}{3\gamma}\frac{c_0^2}{\tan^2{\chi}}\cos{\theta} - \epsilon^2 \frac{2(-1)^N \pi c_0}{3(\gamma-1)}\frac{\sin{\alpha_0}\cos{\chi}}{\sin^3{\chi}}\sin{\psi}\sin{\theta} \nonumber \\
& \qquad -\epsilon^3 \frac{(-1)^N}{3\tan{\chi}} \left(4\cos{\alpha_0} - \cos{3\alpha_0} \right) \sin{\psi}\sin{\theta} - \epsilon^3 \left( 2 + \cos{2\alpha_0} \cos{2\psi} \right)\cos{\theta}.
\label{eqpsi}
\end{align}
Keeping the lowest-order terms and the first corrections, the solution of these equations reads
\begin{align}
& \theta\left(\tilde{t}\right) = \theta_0 - \frac{\epsilon^2}{\omega}\cos{2\alpha_0}\sin{\theta_0}\sin{\epsilon\omega \tilde{t}}\sin{\left(2\psi_0 - \epsilon\omega \tilde{t} \right)}, 
\label{thetaappr}
\\
& \phi\left(\tilde{t}\right)  = \phi_0 + 2\epsilon^3 \tilde{t} + \frac{\epsilon^2}{\omega}\cos{2\alpha_0}\sin{\epsilon\omega \tilde{t}}\cos{\left(2\psi_0 - \epsilon\omega \tilde{t} \right)}, 
\label{phiappr}
\\
& \psi\left(\tilde{t}\right) = \psi_0 - \epsilon \omega \tilde{t} - \frac{\epsilon}{\omega} \frac{2(-1)^N \pi c_0}{3(\gamma-1)}\frac{\sin{\alpha_0}\cos{\chi}}{\sin^3{\chi}}\sin{\theta_0}\big[\cos(\psi_0-\epsilon\omega \tilde{t}) - \cos{\psi_0} \big],
\label{psiappr}
\end{align}
where $\omega=\pi^2 c_0^2 \cos{\theta_0}/(3\gamma \tan^2{\chi})$, and $\theta_0$, $\phi_0$ and $\psi_0$ are the initial values of the corresponding Euler angles. Note, that the terms proportional to $\tilde{t}$ can become arbitrarily large as $\tilde{t}\rightarrow \infty$, independent of their prefactors. Eqs.\eqref{thetaappr}-\eqref{psiappr} show that $\psi$ and $\phi$ grow linearly in time with superimposed small oscillations on top, while $\theta$ oscillates around $\theta_0+(\epsilon^2/2\omega)\cos{2\alpha_0}\sin{\theta_0}\cos{2\psi_0}$ with the amplitude $(\epsilon^2/2\omega)\cos{2\alpha_0}\sin{\theta_0}$.

To verify these predictions, we solve the dimensionless equations of motion numerically. We follow the same methodology as above but keep all the terms in Eqs.\eqref{eom1}-\eqref{eom4}. The resulting equations for the Euler angles are solved numerically using Scientific Python \cite{scipy} by employing the fourth-order Runge-Kutta time-stepping method \cite{canutobook}. 

\begin{figure}[h]
\includegraphics[scale=0.9]{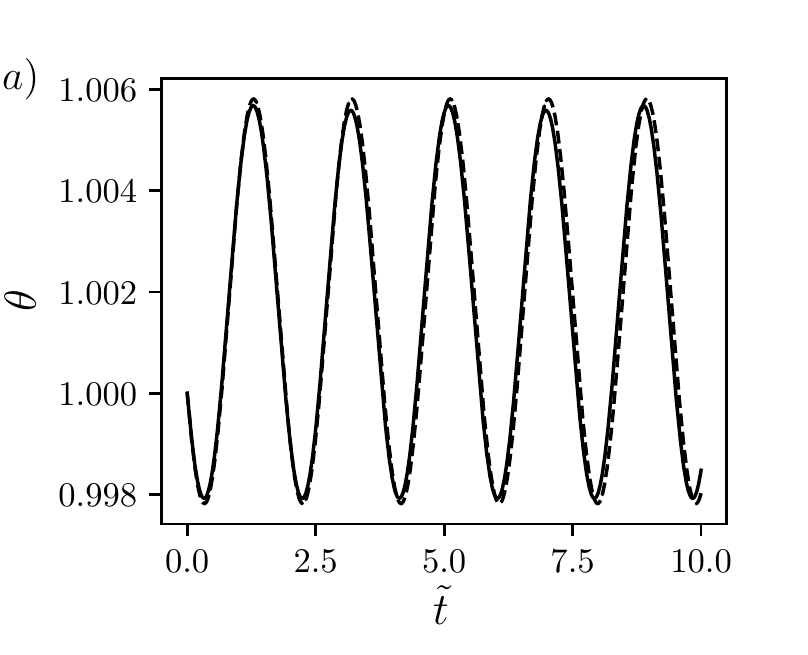}
\includegraphics[scale=0.9]{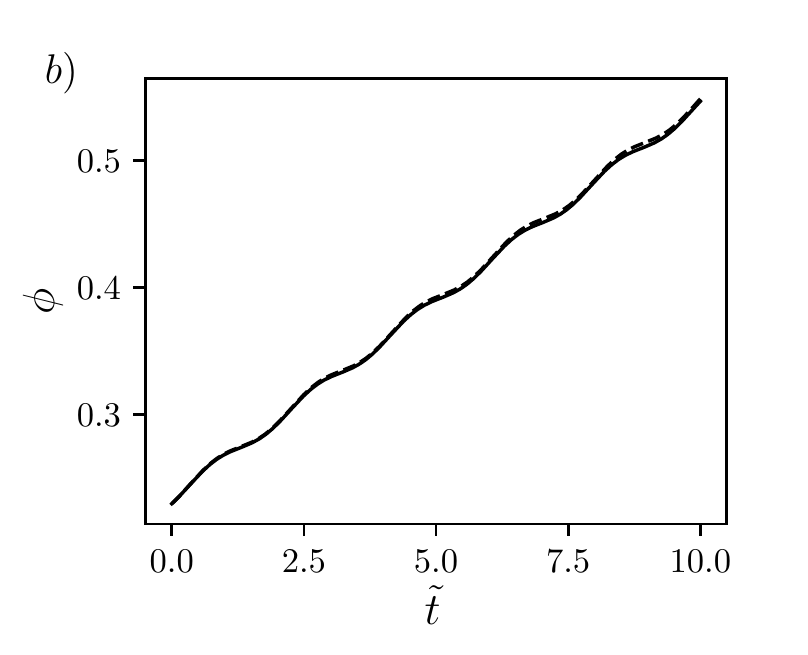}
\includegraphics[scale=0.9]{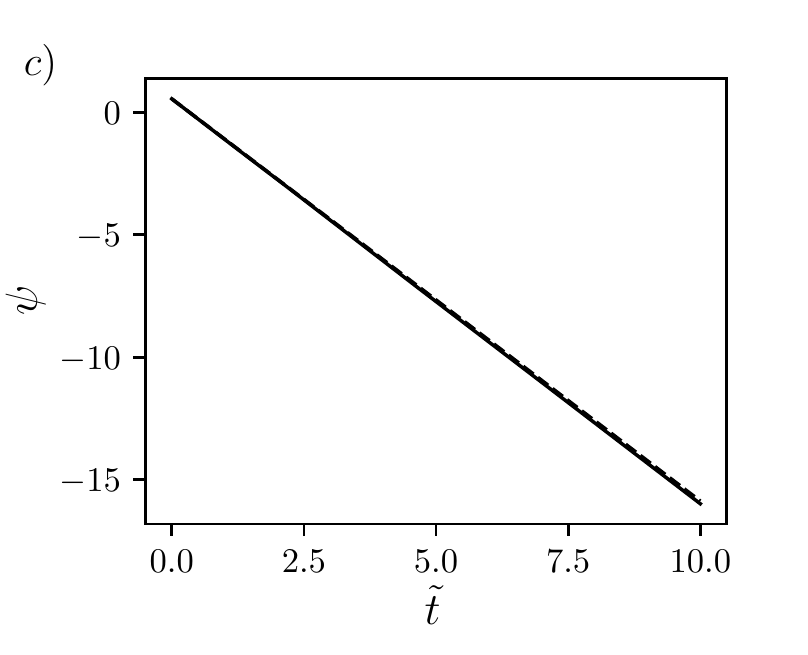}
\caption{Comparison between the predictions of Eqs.\eqref{thetaappr}-\eqref{psiappr} (dashed lines) and the exact numerical solutions (solid lines) of the full equations of motion for a helix with $N = 4$, $\alpha_0=0$, and $\chi = 0.733$ ($\epsilon = 0.25$): a) $\theta(\tilde t)$, b) $\phi(\tilde t)$, and c) $\psi(\tilde t)$. The initial conditions are $\theta_0 = 1.0$, $\phi_0 = 0.2$, and $\psi_0 = 0.5$.}
\label{Euler:good}
\end{figure}

\begin{figure}[h]
\includegraphics[scale=0.9]{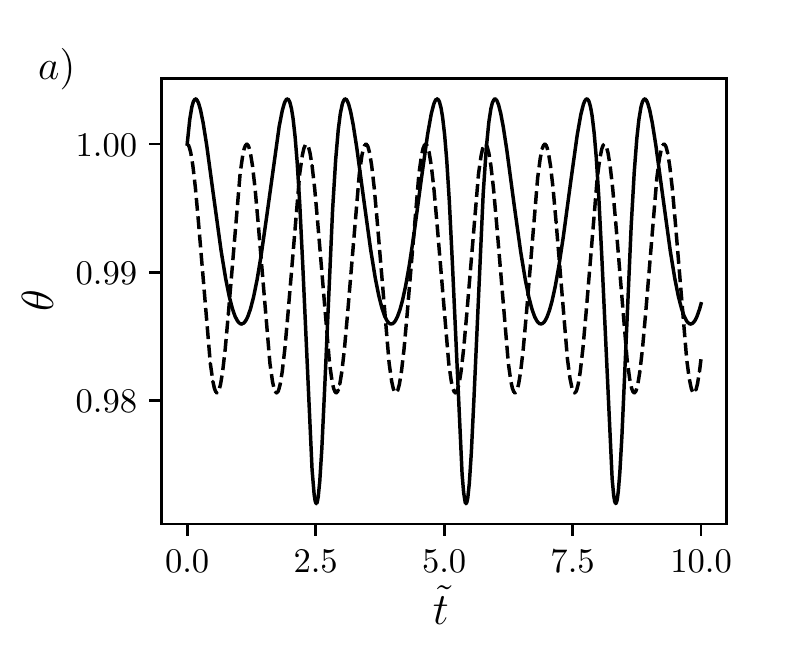}
\includegraphics[scale=0.9]{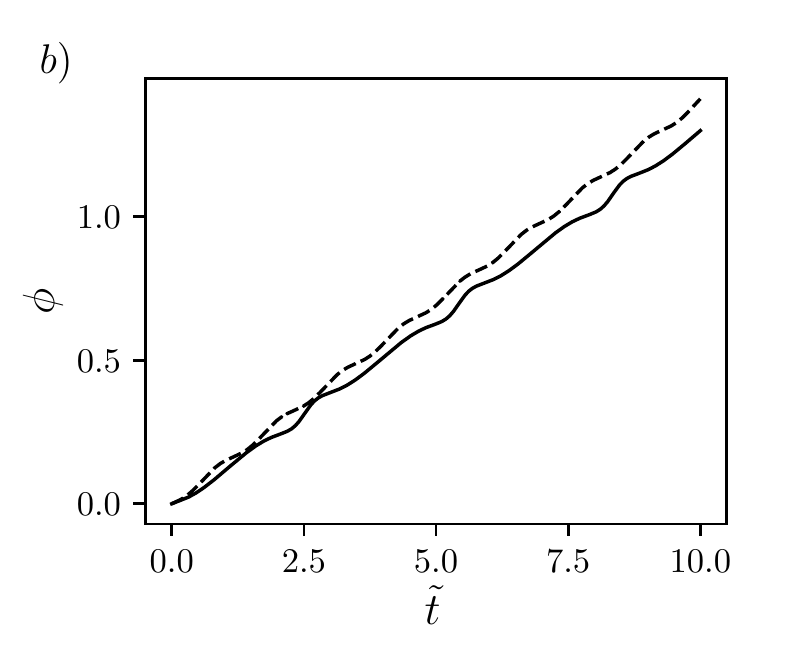}
\includegraphics[scale=0.9]{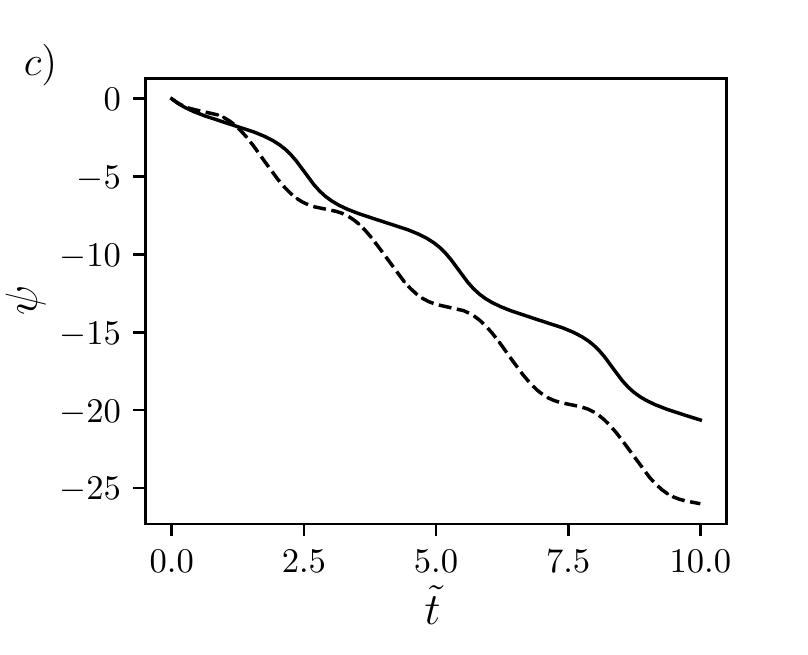}
\caption{Same as Fig.\ref{Euler:good} for a helix with $N = 2$, $\alpha_0=-1.34$, and $\chi = 0.733$ ($\epsilon = 0.41$). The initial conditions are $\theta_0 = 1.0$, $\phi_0 = 0.0$, $\psi_0 = 0.0$.}
\label{Euler:bad}
\end{figure}

In Figs.\ref{Euler:good} and \ref{Euler:bad} we compare the predictions of Eqs.\eqref{thetaappr}-\eqref{psiappr} with the numerical solutions of the full equations. In Fig.\ref{Euler:good} we show the Euler angle dynamics for a relatively long helix with $N=4$, $\alpha_0=0$, and $\chi = 0.733$ (approximately $42$ degrees) starting from arbitrarily chosen initial conditions $\theta_0 = 1.0$, $\phi_0 = 0.2$, and $\psi_0 = 0.5$; solid lines are the exact dynamics, while the dashed lines correspond to Eqs.\eqref{thetaappr}-\eqref{psiappr}. For this geometry $\epsilon=0.25$, and we observe a very good agreement between the numerical solution and the large-$L$ approximation. In Fig.\ref{Euler:bad} we show the Euler angle dynamics for a relatively short helix with $N=2$, $\alpha_0=-1.34$ (approximately $-77$ degrees), and the same $\chi = 0.733$, starting from $\theta_0 = 1.0$, $\phi_0 = 0.0$, and $\psi_0 = 0.0$. This case corresponds to $\epsilon = 0.41$, and appears to exhibit deviations between the exact numerical solution and the large-$L$ limit. We note, however, that even in this case the leading-order prediction of Eqs.\eqref{thetaappr}-\eqref{psiappr} is relatively good: $\theta(\tilde t)$ is well-approximated by a constant value $\theta_0+(\epsilon^2/2\omega)\cos{2\alpha_0}\sin{\theta_0}\cos{2\psi_0}$ (note the scale on the vertical axis of Fig.\ref{Euler:bad}(a), while the slopes of linear increase/decrease of $\phi(\tilde t)$ and $\psi(\tilde t)$ are in reasonable agreement with Eqs.\eqref{phiappr} and \eqref{psiappr}. The main reason why the large-$L$ approximation works relatively well even for short helices is that the leading term in Eq.\eqref{thetaappr} is $O(\epsilon^3)$, which is sufficiently small even for helices with approximately two turns ($\epsilon\sim0.5$). This approximation clearly fails for shorter chiral objects that comprise parts of a single helical turn, which we do not consider here. {\new We remark here that, in general, the accuracy of the large-$L$ approximation is controlled by $\epsilon$. For a fixed $\epsilon$, the magnitude of the discrepancy between the exact numerical solution and the large-$L$ limit prediction is further controlled by $\alpha_0$; these variations, however, are subdominant and vanish in the $L\rightarrow\infty$ limit.}

We now study the implications of Eqs.\eqref{thetaappr}-\eqref{psiappr} for the spatial trajectories of the helix in the lab frame. As discussed above, to leading order the dimensionless equations of motion yield for the velocity components
\begin{align}
&\tilde{U}_1 = \frac{\epsilon^2\pi c_0}{3(\gamma-1)\sin^2{\chi}} \cos{\psi}\sin{\theta}, \\
&\tilde{U}_2= -\frac{\epsilon^2\pi c_0}{3(\gamma-1)\sin^2{\chi}} \sin{\psi}\sin{\theta}, \\
&\tilde{U}_3= \frac{\epsilon^2\pi c_0^2 (\gamma-c_0)}{6\gamma(\gamma-1)\sin^2{\chi}} \cos{\theta}.
\end{align}
Using Eqs.\eqref{TransformRule} and \eqref{RotMatrix}, we transform these velocity components to the shifted lab frame $\{x',y',z'\}$
\begin{align}
&\tilde{U}_{x'} = \frac{1}{2}\left( \frac{\epsilon^2\pi c_0}{3(\gamma-1)\sin^2{\chi}}  - \frac{\epsilon^2\pi c_0^2 (c_0-\gamma)}{6\gamma(\gamma-1)\sin^2{\chi}}\right) \cos{\phi}\sin{2\theta}, \label{Ux}\\
&\tilde{U}_{y'} = \frac{1}{2}\left( \frac{\epsilon^2\pi c_0}{3(\gamma-1)\sin^2{\chi}}  - \frac{\epsilon^2\pi c_0^2 (c_0-\gamma)}{6\gamma(\gamma-1)\sin^2{\chi}}\right) \sin{\phi}\sin{2\theta}, \label{Uy}\\
&\tilde{U}_{z'} = -\frac{\epsilon^2\pi c_0^2 (c_0-\gamma)}{6\gamma(\gamma-1)\sin^2{\chi}}  \cos^2{\theta} - \frac{\epsilon^2\pi c_0}{3(\gamma-1)\sin^2{\chi}}  \sin^2{\theta}. \label{Uz}
\end{align}
Obviously, these components coincide with their values in the $\{X,Y,Z\}$ frame. 
Using the leading order prediction for the Euler angle dynamics, i.e. $\theta\left(\tilde{t}\right) \approx \theta_0$, $\phi\left(\tilde{t}\right)  \approx \phi_0 + 2\epsilon^3 \tilde{t}$, and $\psi\left(\tilde{t}\right) \approx \psi_0 - \epsilon \omega \tilde{t}$, see Eqs.\eqref{thetaappr}-\eqref{psiappr}, the position of the origin of the body frame in the lab frame is given by
\begin{align}
&\tilde{X}(\tilde{t}) = \tilde{X}_0 + \tilde{\rho}\sin{\phi_0} -  \tilde{\rho}\sin{\left(\phi_0+2\pi\frac{\tilde{t}}{\tilde{T}}\right)},  \label{shX} \\
&\tilde{Y}(\tilde{t}) = \tilde{Y}_0 - \tilde{\rho}\cos{\phi_0} +  \tilde{\rho}\cos{\left(\phi_0+2\pi\frac{\tilde{t}}{\tilde{T}}\right)},  \label{shY}  \\
&\tilde{Z}(\tilde{t}) = \tilde{Z}_0 - \tilde{\Lambda}\frac{\tilde{t}}{\tilde{T}},\label{shZ} 
\end{align}
where
\begin{align}
&\tilde\rho = \frac{\pi c_0 \sin{2\theta_0}}{12\epsilon(\gamma-1)\sin^2{\chi}}\left[\frac{c_0(c_0-\gamma)}{2\gamma} - 1 \right], \label{shRho} \\
&\tilde\Lambda = \frac{\pi^2 c_0 }{3\epsilon(\gamma-1)\sin^2{\chi}}\left[ \frac{c_0(c_0-\gamma)}{2\gamma} \cos^2{\theta_0}  +  \sin^2{\theta_0} \right],\label{shLambda}
\end{align}
and the period $\tilde T = \pi/\epsilon^3$; $\left(\tilde{X}_0,\tilde{Y}_0,\tilde{Z}_0\right)$ is its initial position at $\tilde t = 0$. According to Eqs.\eqref{shX}-\eqref{shZ}, the origin of the body frame moves downwards, along the direction of gravity, with the sedimentation speed given by $\tilde{\Lambda}/\tilde T$. In the process, it traces a helical trajectory in the lab frame with the dimensionless radius $|\tilde\rho|$ (note that $\tilde\rho$ can be negative) and pitch $\tilde\Lambda$, and in what  follows, we refer to this trajectory as a \emph{superhelix}. 

\begin{figure}[h]
\includegraphics[width=0.45\textwidth]{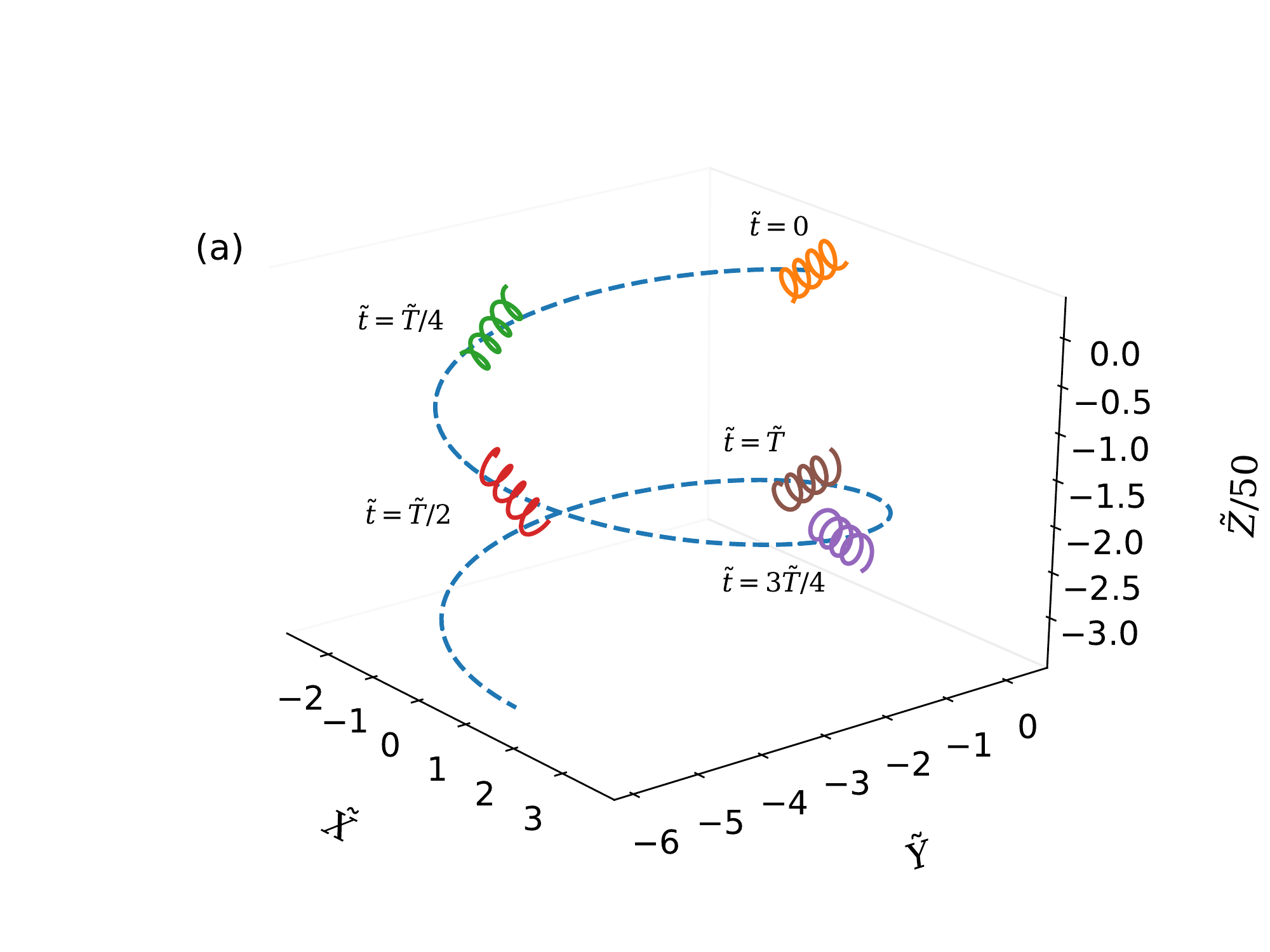}
\hspace{1cm}
\includegraphics[width=0.40\textwidth]{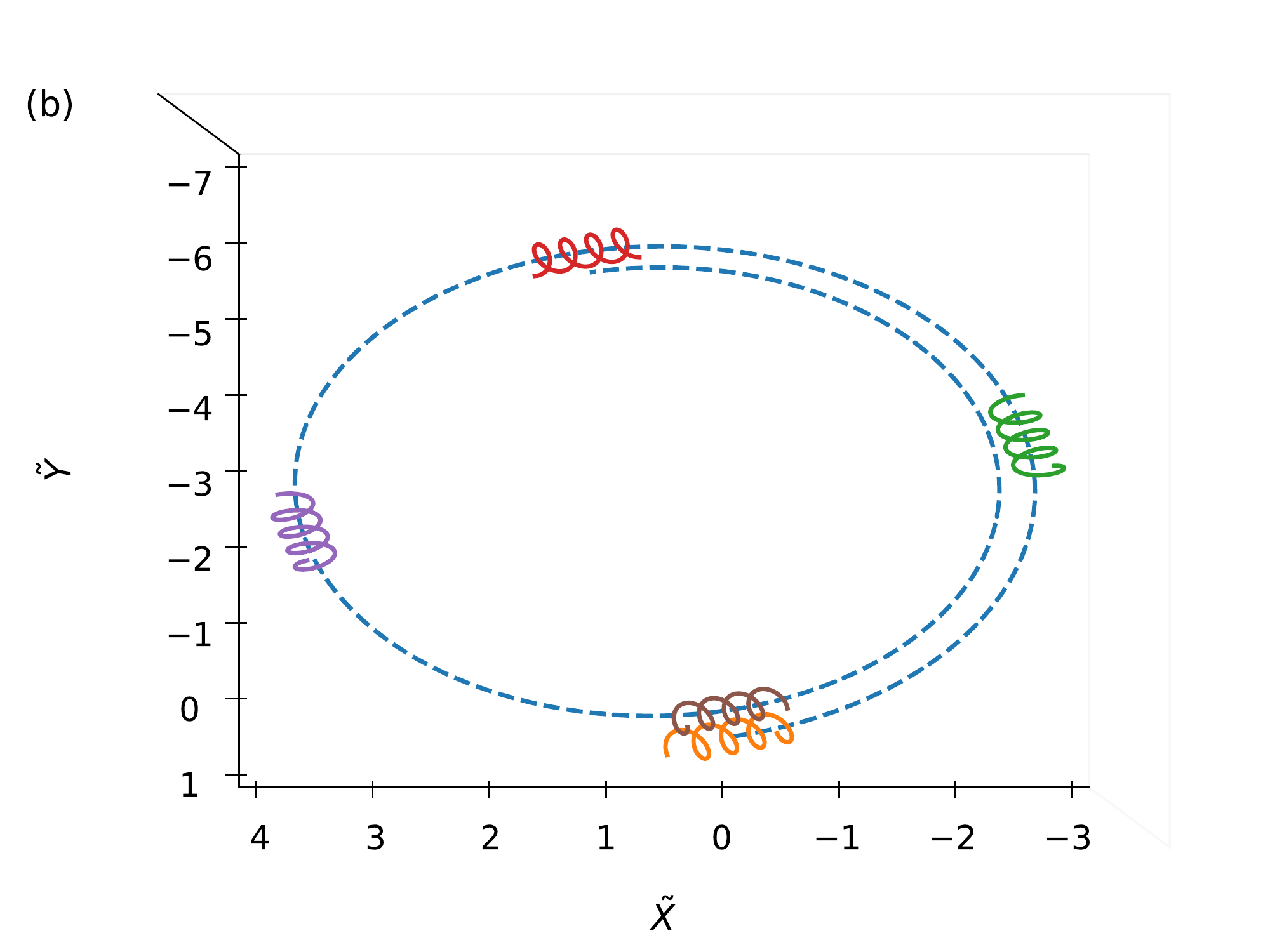}
\caption{Example of a superhelical trajectory traced by the origin of the body frame for a helix with $N = 4$, $\alpha_0=0$,  and $\chi = 0.733$. The initial values of the Euler angles are $\theta_0 = 1.0$, $\phi_0 = 0.2$, and $\psi_0 = 0.5$, while centre of the helix is at $(0,0,0)$ at time $\tilde t = 0$. The dashed line represents the trajectory of the origin of the body frame, given by Eqs.\eqref{shX}-\eqref{shZ}, while colours correspond to the instantaneous orientation of the helix at various times: $\tilde t=0$ (orange), $\tilde t=\tilde{T}/4$ (green), $\tilde t=\tilde{T}/2$ (red), $\tilde t=3\tilde{T}/4$ (violet), and $\tilde t=\tilde{T}$ (brown).  The radius of the helix is increased by a factor of $5$ compared to its actual value for visualisation purposes. a) Side view. b) Top view.}
\label{superhelix:sketch}
\end{figure}

To illustrate the main features of such superhelical trajectories, in Fig.\ref{superhelix:sketch} we plot the analytical predictions for the path traced by the origin of the body frame and the instantaneous orientation for a helix with $N = 4$, $\alpha_0=0$,  and $\chi = 0.733$. The initial values of the Euler angles are $\theta_0 = 1.0$, $\phi_0 = 0.2$, and $\psi_0 = 0.5$, while the origin of the body frame is initially at $(0,0,0)$ in the lab frame. The dashed line represents the trajectory of the centre of the helix, given by Eqs.\eqref{shX}-\eqref{shZ}, while colours correspond to the instantaneous orientation of the helix at various times: $\tilde t=0$ (orange), $\tilde t=\tilde{T}/4$ (red), $\tilde t=\tilde{T}/2$ (green), $\tilde t=3\tilde{T}/4$ (violet), and $\tilde t=\tilde{T}$ (brown). For each time, the orientation of the helix is constructed by applying the rotation matrix, Eq.\eqref{RotMatrix}, to the spatial positions of the material points of the helix, Eq.\eqref{rh}, where the Euler angles are given by Eqs.\eqref{thetaappr}-\eqref{psiappr}. The radius of the helix is made $5$ times larger than its actual value for visualisation purposes. 

First, we observe that both the pitch and the radius of the superhelical trajectory are large, $O(\epsilon^{-1})$, and thus significantly exceed the length of the helix, which is set to unity in our dimensionless units. Also, the pitch is at least ten times larger than the radius of the superhelix (note the difference in scales of the axes) for all helices and the initial values of $\theta_0$, see Eqs.\eqref{shRho} and \eqref{shLambda}.
While the origin of the body frame is moving along the superhelical trajectory, the symmetry axis of the helix, given by ${\bf e}_3 = \left(\cos{\phi}\sin{\theta},\sin{\phi}\sin{\theta},\cos{\theta} \right)$ in the $\{x',y',z'\}$ frame, rotates around $\bf Z$ with the frequency $2\pi/\tilde T$, thus performing exactly one full rotation while travelling down a single pitch $\tilde\Lambda$ of the superhelix. At all times the angle between the symmetry axis of the helix and the tangent to the superhelical trajectory is constant.
Finally, the helix is rotating around its axis of symmetry with $\tilde\Omega_3$, which is much faster than $2\pi/\tilde T$, thus completing $O(\epsilon^{-2})$ turns in one period $\tilde T$. In Fig.\ref{superhelix:sketch} the incommensurate frequencies of both rotations are best observed by comparing the $\tilde t=0$ and $\tilde t=\tilde{T}$ configurations: while the symmetry axis of the helix returns to the same orientation after the full period $\tilde T$, the ends of the helix do not (at $\tilde t=0$ both ends are downwards, while at $\tilde t=\tilde{T}$ they point upwards).

We also note that Eqs.\eqref{shX}-\eqref{shZ} show that the handedness of the superhelical trajectory is opposite to the handedness of the helix: for right-handed helices defined through Eq.\eqref{rh}, trajectories are left-handed helices, while for left-handed helices, defined through Eq.\eqref{rh} with $\lambda\rightarrow -\lambda$, implying $\epsilon \rightarrow -\epsilon$ in the analysis above, Eqs.\eqref{shX}-\eqref{shZ} predict right-handed trajectories. 

\begin{figure}[h]
\includegraphics[scale=0.40]{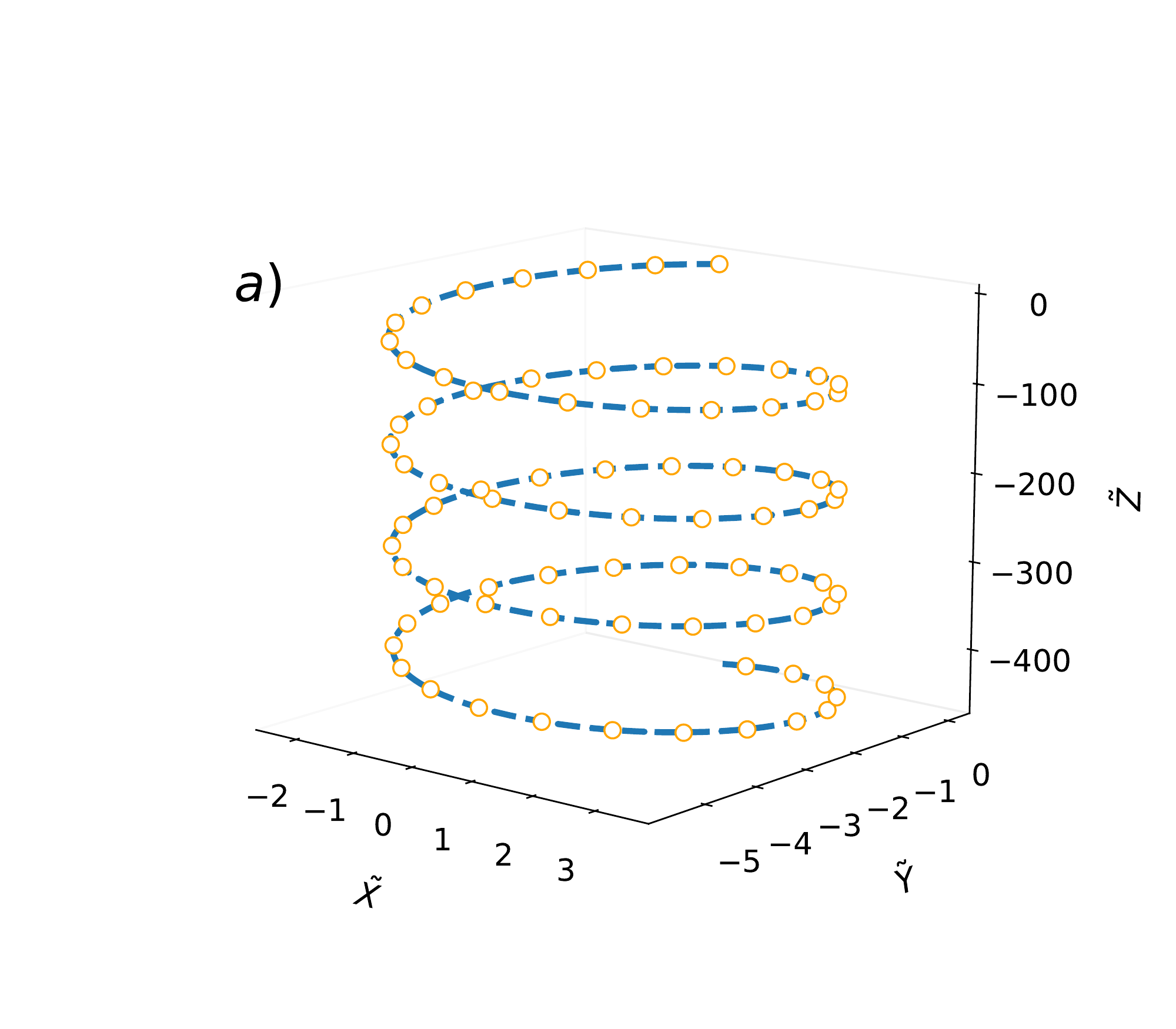}
\includegraphics[scale=0.40]{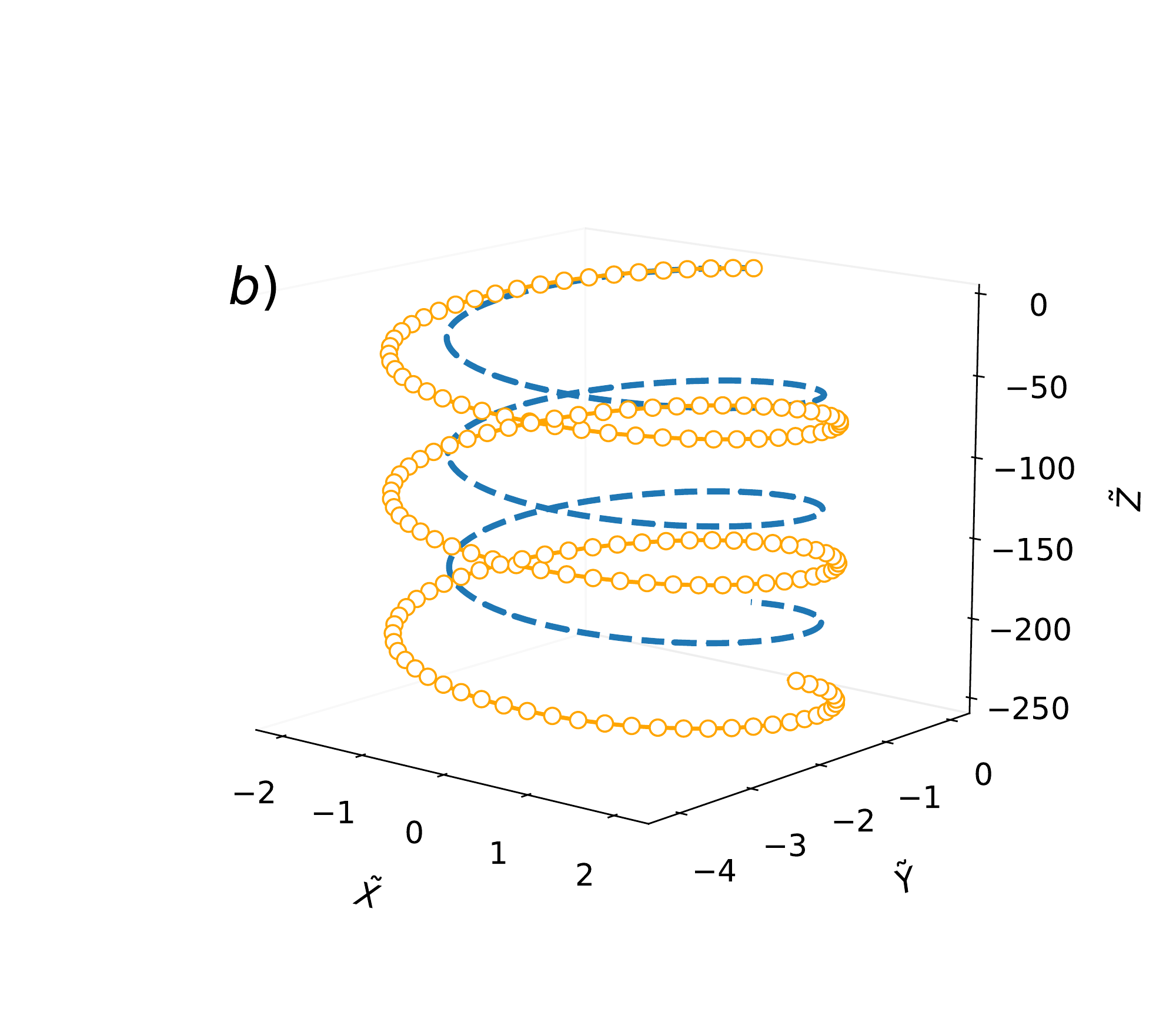}
\caption{Comparison between the superhelical trajectories predicted by Eqs.\eqref{shX}-\eqref{shZ} (dashed lines) and the numerical solutions of the full equations (circles and solid lines). a) Long helix: $N=4$, $\alpha_0=0$, and $\chi = 0.733$ with $\theta_0 = 1.0$, $\phi_0 = 0.2$, and $\psi_0 = 0.5$. b) Short helix: $N=2$, $\alpha_0=-1.34$, and $\chi = 0.733$, with $\theta_0 = 1.0$, $\phi_0 = 0.0$, and $\psi_0 = 0.0$.}
\label{superhelix:comp}
\end{figure}

In Fig.\ref{superhelix:comp} we compare the superhelical trajectories predicted by Eqs.\eqref{shX}-\eqref{shZ} against the numerical solution of the full equations. As above, we study two cases: a relatively long helix with $N=4$, $\alpha_0=0$, and $\chi = 0.733$ starting from $\theta_0 = 1.0$, $\phi_0 = 0.2$, and $\psi_0 = 0.5$, shown in Fig.\ref{superhelix:comp}a), and a short helix with $N=2$, $\alpha_0=-1.34$, and $\chi = 0.733$, starting from $\theta_0 = 1.0$, $\phi_0 = 0.0$, and $\psi_0 = 0.0$, shown in Fig.\ref{superhelix:comp}b). The dashed lines are the large-$L$ predictions, while the open circles are the numerical solution. As with the dynamics of the Euler angles, the large-$L$ prediction is only qualitatively correct for the short helix, but shows excellent agreement for the longer one. 

According to Eq.\eqref{shRho}, the analytical approximation predicts that the radius of the superhelical trajectory changes sign at the value of $\chi$ given by $\sin^2{\chi_0}=4/(3+\sqrt{(8+\gamma)/\gamma})$, which for $\gamma=2$ implies $\chi_0=1.06$ (about $61$ degrees). We note that the sign of $\tilde\rho$ does not have an effect on the handedness of the superhelical trajectories, and the conclusion reached above holds on both sides of this critical value. At $\chi=\chi_0$, the analytical theory breaks down, and higher-order terms should be retained in Eqs.\eqref{eqtheta}-\eqref{eqpsi}. We did not explore this regime as we do not expect the resistive force theory to be sufficiently accurate at such high values of $\chi$ \cite{Pak2012,Jung2007,Rodenborn2013,Koens2014}.

Finally, we note that there are two special orientations, corresponding to the vertical and horizontal sedimentation, that should be discussed separately. The former is given by the $\theta_0\rightarrow 0$ limit of Eqs.\eqref{shX}-\eqref{shZ}, and requires no special treatment. The latter, however, requires
additional analysis, as we show in the next Section.

\section{Sedimentation in almost-horizontal orientations}
\label{Section:horizontal}

The large-$L$ approximation for the dynamics of the Euler angles, Eqs.\eqref{thetaappr}-\eqref{psiappr}, developed in the previous Section, relies on the assumption that the first term on the r.h.s. of Eq.\eqref{eqpsi}, $O(\epsilon)$, is the dominant one. In an almost-horizontal orientation, when the initial condition $\theta_0$ is close to $\pi/2$, the first term on the r.h.s. of Eq.\eqref{eqpsi} becomes small compared to the second term, even though it is $O\left(\epsilon\right)$, while the second term is $O\left(\epsilon^2\right)$. Comparing the two terms yields the following estimate for the transitional value of $\theta_0$, where the $O\left(\epsilon^2\right)$-contribution in Eq.\eqref{psiappr} becomes the dominant term
\begin{align}
\theta_0^{(tr)} \approx \frac{\pi}{2} - \Bigl\lvert \frac{4\epsilon\gamma}{\pi c_0 (\gamma-1)}\frac{\sin{\alpha_0}}{\sin{2\chi}} \Bigl \rvert .
\label{theta_tr}
\end{align}
When $\alpha_0$ is close to zero, corresponding to helices with almost-integer number of turns, the dominant term in Eq.\eqref{psiappr} is the $O\left(\epsilon^3\right)$-contribution. However, the associated range of $\alpha_0$ is so small that we do not discuss it here. For $\theta_0>\theta_0^{(tr)}$, the solution to Eqs.\eqref{eqtheta}-\eqref{eqpsi} changes significantly. Before discussing the dynamics of the Euler angles and the corresponding spatial trajectories for this regime, we analyse the linear stability of the strictly horizontal orientation.

\begin{figure}[h]
\includegraphics[scale=0.40]{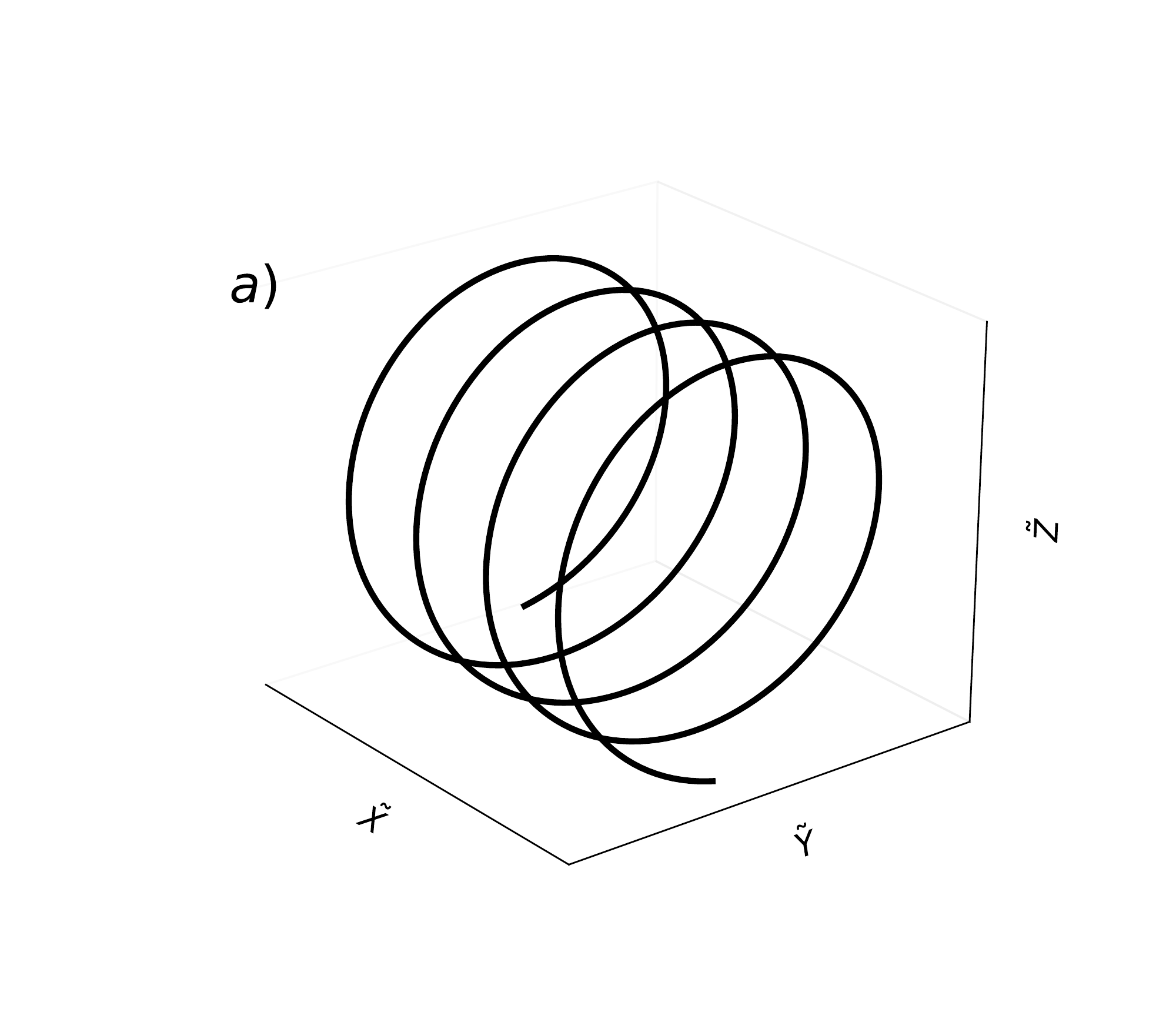}
\includegraphics[scale=0.40]{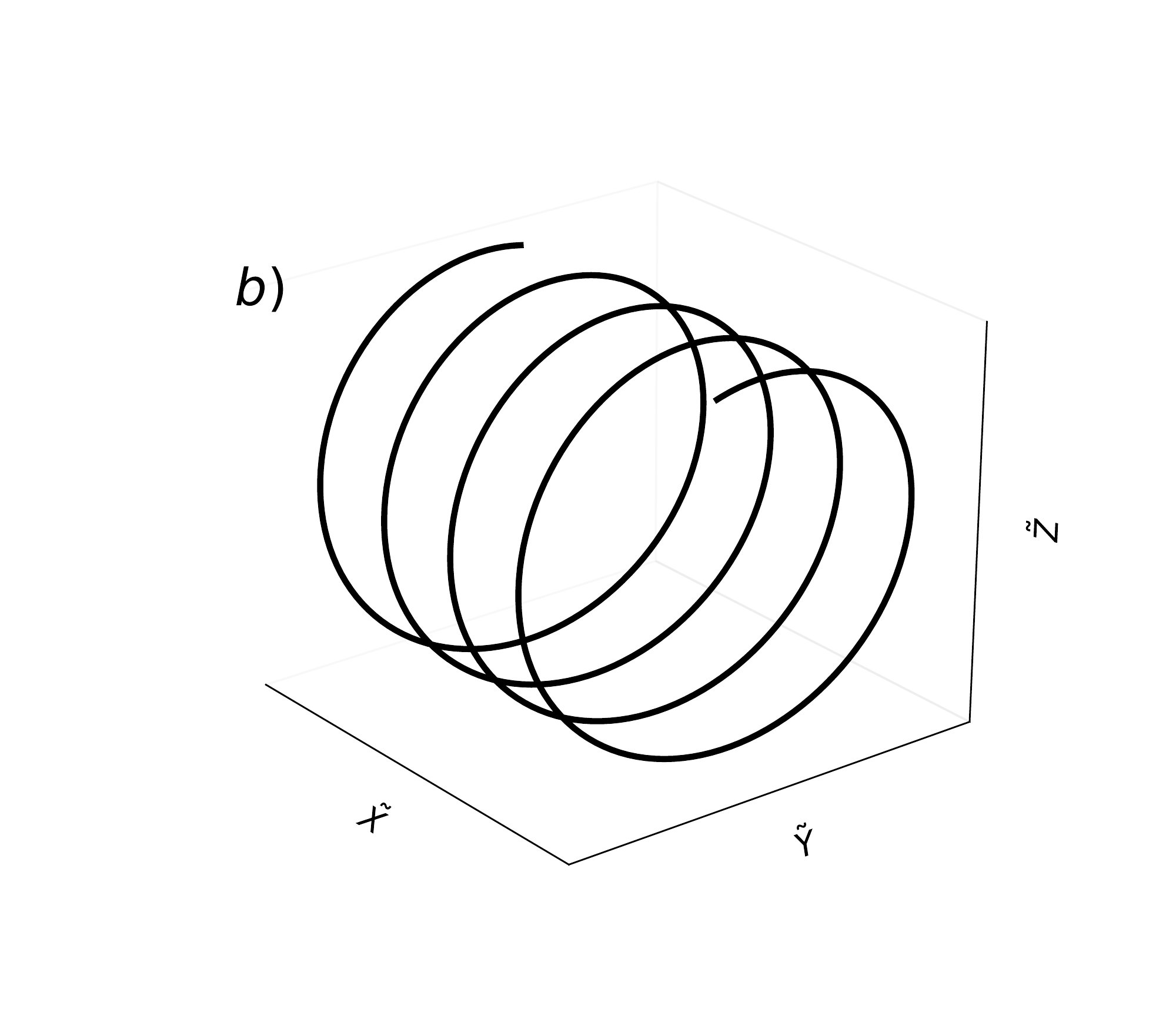}
\caption{Examples of stable horizontal configurations for a helix with $N=4$: a) $\alpha_0=0.24$, the stable configuration is given by $\psi=0$. b) $\alpha_0=-0.24$, the stable configuration is given by $\psi=\pm\pi$ (both configurations look the same). The relative orientation in the $\tilde{X}\tilde{Y}$-plane is chosen arbitrarily as the helix is rotating around the $\tilde{Z}$ axis.}
\label{horizontal:orientation}
\end{figure}

When $\theta_0=\pi/2$, the r.h.s. of Eq.\eqref{eqtheta} vanishes, implying $\theta(\tilde t)=\theta_0$, while the evolution of $\psi$ is given by the leading term in Eq.\eqref{eqpsi}
\begin{align}
\frac{\partial \psi}{\partial \tilde t} = - A \sin{\psi},
\end{align}
where
\begin{align}
A = \epsilon^2 \frac{2(-1)^N \pi c_0}{3(\gamma-1)}\frac{\sin{\alpha_0}\cos{\chi}}{\sin^3{\chi}}.
\end{align}
The solution to this equation,
\begin{align}
\tan\left({\frac{\psi(\tilde t)}{2}}\right) = \tan{\left(\frac{\psi_0}{2}\right)}\,e^{-A\,\tilde t},
\end{align}
demonstrates that $\psi$ asymptotically approaches a constant value, determined by the sign of the constant $A$, which, in turn, is set by the sign of $(-1)^N \sin{\alpha_0}$: for even $N$ and $\alpha_0>0$, and for odd $N$ and $\alpha_0<0$, $\psi\rightarrow 0$ as $\tilde t \rightarrow\infty$, while for even $N$ and $\alpha_0<0$, and for odd $N$ and $\alpha_0>0$, $\psi\rightarrow \pm\pi$, depending on the initial value $\psi_0$. Based on our parametrisation of the helix, Eq.\eqref{rh} and the discussion after it, we can translate these rather abstract statements into a simple geometrical interpretation: helices with a number of turns smaller than the closest  integer $N$ (i.e. helices with $\alpha_0 > 0$) sediment with their free ends pointing downwards, while helices with the number of turns larger than the closest integer ($\alpha_0 < 0$) sediment with their free ends pointing upwards, see Fig.\ref{horizontal:orientation}, for example. As can be seen from Eqs.\eqref{eqtheta}-\eqref{eqpsi}, $\theta=\pi/2$ and $\psi=0$ or $\psi = \pm \pi$ are stationary points of those equations, while $\phi$ increases linearly in time, $\phi\left(\tilde t\right)  = \phi_0 + 2\epsilon^3 \tilde t$. This implies that horizontally oriented helices move in a straight line along the direction of gravity with the velocity $\tilde{U}_{z'}$ given by Eq.\eqref{Uz} with $\theta=\pi/2$, while rotating around the vertical axis with the dimensionless angular velocity $2\epsilon^3$.

To study the linear stability of the horizontal orientation, we consider a small perturbation to the Euler angles, $\theta(\tilde t)=\pi/2-\delta\theta(\tilde t)$ and $\psi(\tilde t)=\psi_0+\delta\psi(\tilde t)$, where $\psi_0$ is either $0$ or $\pm \pi$, see above. Assuming that $\delta\theta$ and $\delta\psi$ are infinitesimal, we linearise Eqs.\eqref{eqtheta} and \eqref{eqpsi}, and obtain the following equation for the perturbation
\begin{align}
\frac{\partial^2\delta\theta}{\partial \tilde{t}^2} = - \epsilon^4 \frac{\pi^2 c_0^2 \cos{2\alpha_0}}{3\gamma\tan^2{\chi}}  \delta\theta.
\end{align}
This equation has exponentially growing solutions for $|\alpha_0|>\pi/4$, and we, therefore, conclude that the horizontal orientation is stable with respect to small perturbations for helices with $|\alpha_0|<\pi/4$, and is unstable, otherwise.
As a result, the dynamics of the Euler angles for $\theta_0>\theta_0^{(tr)}$ depend strongly on the value of $\alpha_0$, as we now demonstrate.

\begin{figure}[h]
\includegraphics[scale=0.9]{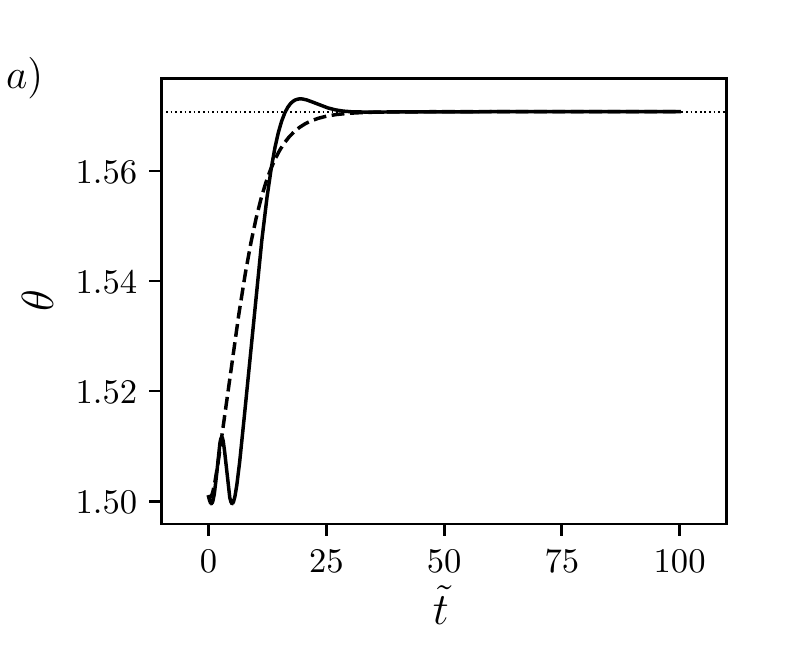}
\includegraphics[scale=0.9]{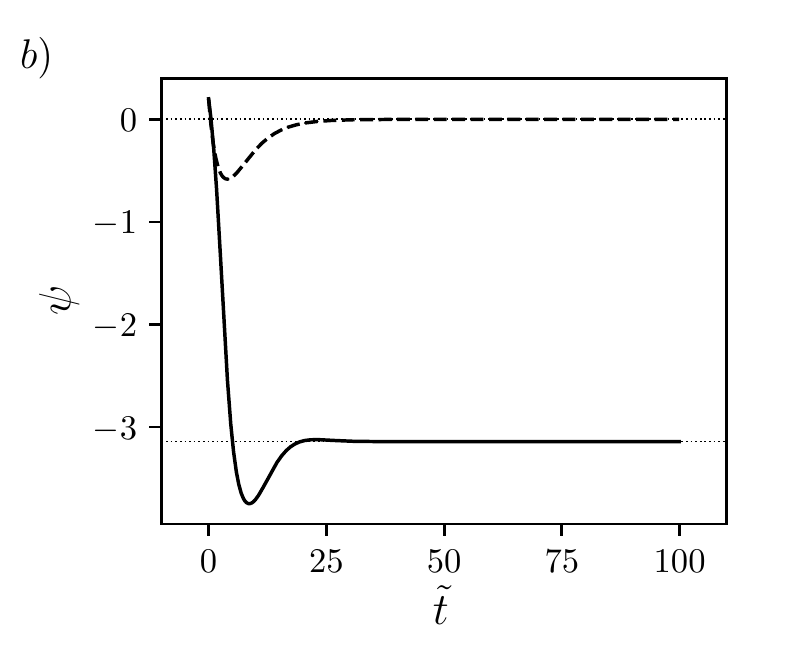}
\caption{Numerical solution of the full equations of motion for a) $\theta(\tilde t)$ and b) $\psi(\tilde t)$. The parameters of the helix are $N=4$ and $\chi = 0.733$, and the initial conditions are given by $\theta_0 = \pi/2 - 0.07$, $\phi_0 = 0.2$, and $\psi_0 = 0.2$. The solid and dashed lines correspond to $\alpha_0=-0.5$ and $\alpha_0=0.5$, respectively. In a) the dotted line is at $\theta=\pi/2$. In b) the dotted lines are at $\psi=0$ and $\psi=-\pi$.}
\label{smallTsteady}
\end{figure}

In Fig.\ref{smallTsteady}, we plot $\theta(\tilde t)$ and $\psi(\tilde t)$ obtained by numerically solving the full equations of motion for a helix with $N=4$ and $\chi = 0.733$, with $\theta_0 = \pi/2 - 0.07$, $\phi_0 = 0.2$, and $\psi_0 = 0.2$ being the initial values for the corresponding Euler angles; the solid lines in Fig.\ref{smallTsteady} correspond to $\alpha_0=-0.5$, while the dashed lines correspond to $\alpha_0=0.5$. This combination of the helix parameters and the initial values corresponds to the regime $\theta_0>\theta_0^{(tr)}$, discussed above ({\new $\theta_0^{(tr)} \approx \pi/2 - 0.083$ }for this case). As Fig.\ref{smallTsteady} indicates, in this regime the dynamics are attracted towards the horizontal orientation, $\theta=\pi/2$, which is linearly stable for $\alpha_0=\pm 0.5$. When $\alpha_0=-0.5$, $\psi$ approaches $-\pi$, while for  $\alpha_0=0.5$, $\psi$ goes to zero, in line with the discussion of stable horizontal orientations above.

\begin{figure}[h]
\includegraphics[scale=0.40]{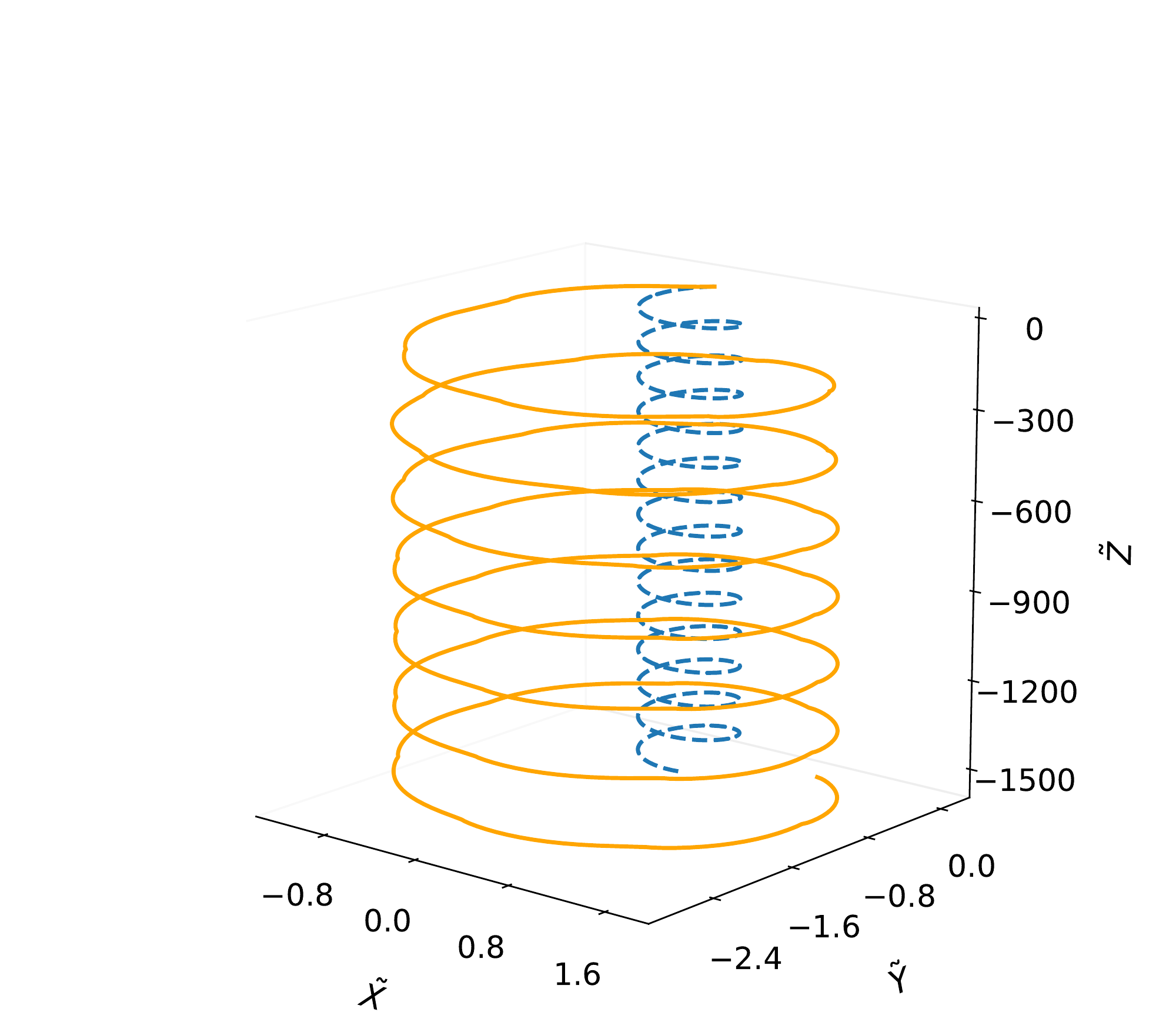}
\includegraphics[scale=0.7]{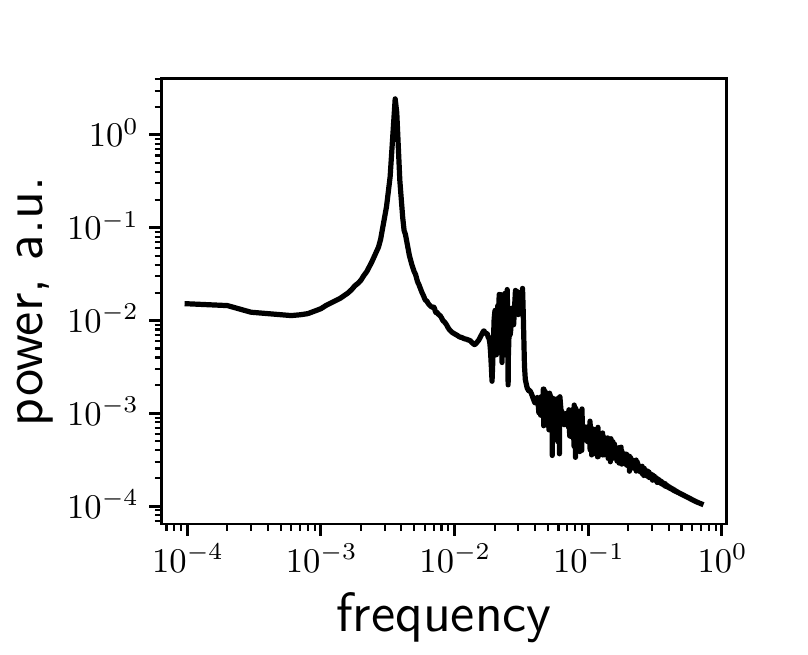}
\caption{a) Spatial trajectory obtained by numerical integration of the exact equations of motion (solid line) for a helix with $N = 4$, $\alpha_0=-1.3$ and $\chi = 0.733$ ($\epsilon = 0.226562$) for $\theta_0 = \pi/2-0.05$, $\phi_0 = 0.2$, and $\psi_0 = 0.2$. The analytical superhelical trajectory predicted by Eqs.\eqref{shX}-\eqref{shZ}, dashed line, is given for reference. b) Power spectrum of  $\tilde{X}(\tilde t)$ corresponding to the trajectory in a).}
\label{unsteady}
\end{figure}

This behaviour changes significantly when $\lvert\alpha\rvert>\pi/4$. In this regime, a trajectory starting from $\theta_0>\theta_0^{(tr)}$ is still being attracted towards the horizontal configuration, but the latter is now linearly unstable, the trajectory is pushed away from the horizontal orientation, and the whole process repeats itself, leading to (quasi-)periodic oscillations of $\theta(\tilde t)$. These oscillations are similar to the oscillations of $\theta(\tilde t)$ in the superhelical regime albeit with a significantly larger amplitude: while the amplitude of $\theta$ oscillations in the superhelical regime is $O(\epsilon^2)$, as can be seen from Eq.\eqref{thetaappr}, in this regime the amplitude is controlled by the size of the basin of attraction of the horizontal orientation, i.e. $\pi/2 - \theta_0^{(tr)}\sim O(\epsilon)$. In Fig.\ref{unsteady}a) we plot the spatial trajectory traced by the origin of the body frame in the lab frame (solid lines) obtained by numerical integration of the full equations of motion for a helix with $N=4$, $\alpha_0=-1.3$, $\chi = 0.733$, where the initial conditions are $\theta_0 = \pi/2 - 0.05$, $\phi_0 = 0.2$, and $\psi_0 = 0.2$. For reference, we also plot the superhelical trajectory (dashed line), Eqs.\eqref{shX}-\eqref{shZ}, for the same values of the parameters. As can be seen from Fig.\ref{unsteady}a), the oscillations in $\theta$ result in a superhelical-like spatial trajectory, although its characteristics are no longer given by Eqs.\eqref{shX}-\eqref{shZ}. The observed radius and the pitch of the trajectory are significantly larger than their superhelical counterparts, Eqs.\eqref{shRho} and \eqref{shLambda}, and the trajectory appears to be less regular. To assess this irregularity, in Fig.\ref{unsteady}b) we plot the power spectrum of  $\tilde{X}(\tilde t)$ for the trajectory in Fig.\ref{unsteady}a). While the main peak, associated with the superhelical component of the trajectory, is still prominent in the power spectrum, there are many other frequencies involved, although the dynamics do not seem to be chaotic. Very long-time numerical solutions (up to $\tilde t \sim 10^7$, not shown) suggest that the radius of this quasi-superhelical trajectory is slowly decreasing until it reaches a steady-state, although we cannot be sure whether this is the true behaviour of the system or whether it is caused by accumulation of numerical errors when solving very non-linear equations for the Euler angles \cite{Hinch1979}. In any case, the very long timescales associated with these changes make them likely to be irrelevant in practice.

\begin{figure}[h]
\includegraphics[scale=0.65]{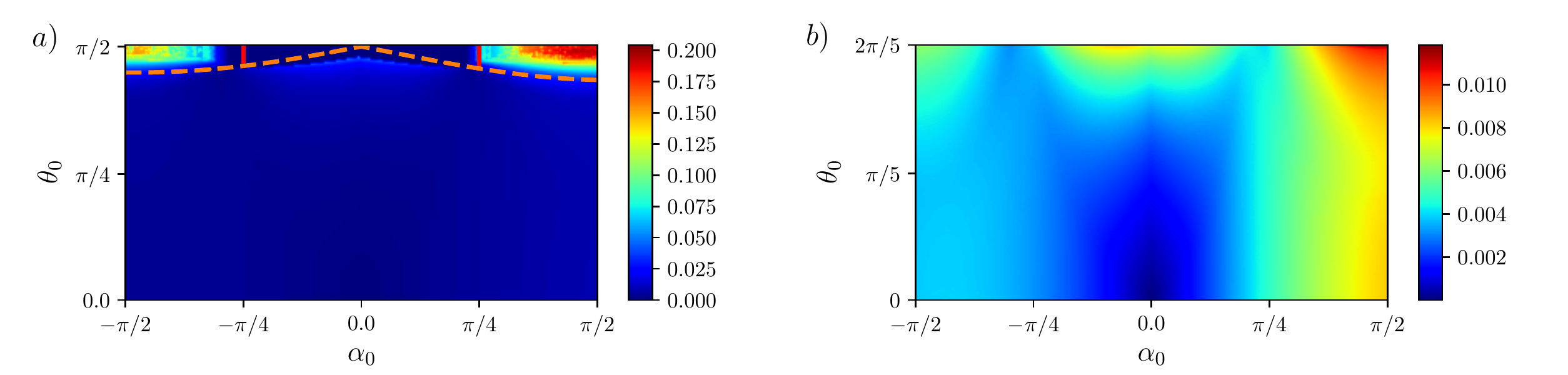}
\caption{State diagram for a helix with $N=4$ and $\chi = 0.733$. For a given combination of $\alpha_0$ and $\theta_0$, the exact equations of motion are numerically integrated with $\phi_0=\psi_0=0$. The colour indicates the amplitude of oscillations of $\theta(\tilde t)$ in a steady state or at long times, if the steady state is not reached, see the main text. a) The dashed line gives the analytical prediction for the boundary between the superhelical and horizontal orientations $\theta_0^{(tr)}$, Eq.\eqref{theta_tr}. The solid lines are the analytical predictions, $\alpha_0=\pm\pi/4$, for the threshold between the linearly stable horizontal orientations and quasi-superhelical trajectories with large oscillations of $\theta(\tilde t)$. b) Zoom-in for $\theta_0\lesssim\theta_0^{(tr)}$.}
\label{PhaseDiag}
\end{figure}

\section{Discussion}

We summarise our findings in Fig.\ref{PhaseDiag} using a representative example of a helix with $N=4$ and $\chi = 0.733$. As discussed in the previous Sections, the type of the spatial trajectory can be inferred from the dynamics of the Euler angle $\theta$, and we use this property to delineate the parameter space. We numerically solve the full equations of motion for the Euler angles starting from $\phi_0=\psi_0=0$, while $\theta_0$ is varied in the range $\left[0,\pi/2\right]$; the geometric parameter $\alpha_0$, that controls how close is the number of pitches in the helix to an integer, see Eq.\eqref{rh}, is varied within the range $\left[-\pi/2,\pi/2\right]$. For each set of parameters, we measure the amplitude $\Delta$ of oscillations of $\theta$ either when the dynamics have reached a steady state, or after three full periods of oscillation, when a true steady state is not achieved (as in Fig.\ref{unsteady}). The results are plotted in Fig.\ref{PhaseDiag} as a function of $\alpha_0$ and the initial value $\theta_0$.

The largest region in Fig.\ref{PhaseDiag} is occupied by weak oscillations in $\theta$ close to the initial $\theta_0$, associated with the superhelical solution, Eqs.\eqref{thetaappr}-\eqref{psiappr} and \eqref{shX}-\eqref{shZ}. This behaviour changes above some critical value $\theta_0^{(tr)}$, and we observe that the estimate of $\theta_0^{(tr)}$, Eq.\eqref{theta_tr}, (dashed line in Fig.\ref{PhaseDiag}), is in good agreement with the numerical data as long as $\alpha_0$ is not too small, where it underestimates the critical value. For $\theta_0>\theta_0^{(tr)}$, the helix either ends up in a linearly stable horizontal orientation, when $\lvert\alpha_0\rvert<\pi/4$, or follows an irregular superhelix-like spatial trajectory with a very large radius and pitch, when $\lvert\alpha_0\rvert>\pi/4$, and the horizontal orientation is predicted to be linearly unstable. The transition thresholds, $\lvert\alpha_0\rvert=\pi/4$, are given by solid lines in Fig.\ref{PhaseDiag}. Note, that while this estimate works rather well for positive values of $\alpha_0$, there are some discrepancies for $\alpha_0<0$, underlying the approximate nature of the linear stability analysis of the horizontal orientation from the previous Section.

The superhelical trajectories that we predict for a wide range of parameters can be understood in a rather intuitive way. Since the helix is an elongated object, it is expected to sediment with an instantaneous velocity lying in the plane spanned by its axis of symmetry and the direction of gravity
and forming a non-zero angle to both direction \cite{HappelBrenner}. The chirality of the helix ensures that this motion then causes the helix to rotate around the direction of gravity and around its axis of symmetry, due to the translational-rotational coupling \cite{HappelBrenner}, leading to a steady rotation of the sedimentation plane mentioned above. The resulting path traced by the origin of the body frame in the lab frame is a superhelix. 

Our calculations also predict the dynamics of a helix in two limiting cases: sedimentation in the vertical and horizontal orientations. In both cases the helix moves in a straight path along the direction of gravity and simultaneously rotates around it. For the vertical orientation, the sedimentation speed is given by Eq.\eqref{Uz} with $\theta=0$, while the rotation rate around its axis of symmetry is given by $\tilde\Omega_{z'} = \dot\phi + \dot\psi = -\epsilon \pi^2 c_0^2/(3\gamma \tan^2{\chi})$,
where we used Eqs.\eqref{eqphi} and \eqref{eqpsi}. The sedimentation speed is $O(\epsilon^2)$, while the rotation rate is $O(\epsilon^{-1})$, and in one period of rotation the helix travels the dimensionless distance $\epsilon (c_0-\gamma)/\left((\gamma-1)\cos^2{\chi}\right)$, which is a small portion of the length of the helix. For the horizontal orientation, the sedimentation speed is given by Eq.\eqref{Uz} with $\theta=\pi/2$, while the rotation rate around the direction of gravity is $\tilde\Omega_{z'} = \dot\phi=2\epsilon^3$. In this case the distance travelled in one period of rotation is significantly larger than the length of the helix, $\pi^2 c_0/( 3\epsilon (\gamma-1) \sin^2{\chi} )>1$. For helices with $\lvert \alpha_0\rvert <\pi/4$, the stable horizontal orientation corresponds to its free ends pointing downwards (along the direction of gravity) when $\alpha_0>0$, and to its ends pointing upwards, for $\alpha_0<0$; we predict no stable horizontal orientation for helices with $\lvert \alpha_0\rvert >\pi/4$. The ratio of the vertical to horizontal sedimentation speeds, $c_0(c_0-\gamma)/(2\gamma)$, which reduces to $\gamma$ for a straight rod, $\chi=0$, may provide a suitable quantity to compare our resistive-force theory against experiments or more accurate theoretical approaches.

It is rather unlikely that the superhelical trajectories can be observed experimentally with macroscopic helices in confined geometries, like a fluid in a tank. According to Eq.\eqref{shRho}, the radius of the superhelical trajectory in physical units, $L \tilde \Lambda$, scales as $L^2/\lambda$, which implies a very wide trajectory. In turn, the size of the tank used in such an experiment should be significantly larger than $L \tilde \Lambda$, implying rather wide geometries. Worse still, the pitch of the superhelical trajectory is significantly larger than its radius, implying not only wide, but also very tall fluid tanks. For instance, consider a helix with $L=10$cm, $\chi=0.733$ and four full turns, $N=4$ and $\alpha_0=0$. If the radius of the helical filament is, say, $r=0.5$mm, the ratio of the friction coefficients becomes $\gamma \approx 1.26$, where we used Eqs.\eqref{Kpar} and \eqref{Kperp} for $K_{\parallel}$ and $K_{\perp}$. With these parameters, and selecting $\theta_0=\pi/4$, we obtain $ \rho \approx 2 L=20$cm and $\Lambda \approx 285 L=28.5$m! For a copper filament, $\rho_h=8.96$g/cm$^3$, suspended in glycerol at room temperature with $\rho_f=1.26$g/cm$^3$ and $\mu=1.41$Pa$\cdot$s \cite{Cheng2008}, the sedimentation velocity, given by Eq.\eqref{Uz}, becomes $L \tilde{U}_{z'}/\tau=7$mm/s.

Recent advances in manufacturing \cite{Attia2009,Pham2013} and manipulation of microhelices under flow conditions \cite{Attia2009,Pham2015} suggest that it could be more appropriate to look for microfluidic realisations of superhelical trajectories. Due to the linear nature of the Stokes equation and the absence of a lengthscale in the problem besides the dimensions of the helix, all lengths in the estimates above will still be correct when scaled by a common factor. Therefore, for $50\mu$m-long helices, which is within the range used by Pham \emph{et al.} \cite{Pham2015}, for instance, one would need a $1$cm-long microfluidic channel to detect superhelical trajectories.

In this work we employed the resistive-force theory \cite{Lighthill73} to approximate the drag on a section of helical filament. As mentioned in Introduction, at best, the resistive-force theory produces semi-quantitative approximations to actual drag forces on extended objects \cite{Jung2007,Rodenborn2013,Koens2014}, while, at worst, it fails even qualitatively for compact objects where hydrodynamic interactions between remote parts of the object are crucial \cite{Pak2012}.
Therefore, our results should be seen as indicative and more detailed studies are necessary to verify their range of applicability. Recently, some preliminary results were obtained numerically for sedimentation of helices within the slender-body theory \cite{Hassan2015}. The spatial trajectory reported in \cite{Hassan2015} appears to fluctuate randomly in the regime where we predict periodic motion associated with the superhelical trajectory, and further work is required to find the origin of this discrepancy.

\bibliography{lit}

\end{document}